\documentclass[a4paper,11pt]{article}
\pdfoutput=1
\usepackage{jheppub}
\usepackage{amsmath}
\usepackage{multirow}
\newcommand{\ii}{\mathrm{i}}

\newcommand{\Tr}{\mathrm{Tr}\,}
\newcommand{\tr}{\mathrm{tr}\,}

\newcommand{\cA}{{\mathcal{A}}}

\newcommand{\cN}{{\mathcal{N}}}
\newcommand{\cO}{{\mathcal{O}}}
\newcommand{\cP}{{\mathcal{P}}}

\newcommand{\cS}{{\mathcal{S}}}

\newcommand{\cW}{{\mathcal{W}}}
\newcommand{\cX}{{\mathcal{X}}}
\newcommand{\cY}{{\mathcal{Y}}}
\newcommand{\cZ}{{\mathcal{Z}}}

\newcommand{\one}{{\rm 1\kern -.9mm l}}

\newdimen\tableauside\tableauside=1.0ex
\newdimen\tableaurule\tableaurule=0.4pt
\newdimen\tableaustep
\def\phantomhrule#1{\hbox{\vbox to0pt{\hrule height\tableaurule
width#1\vss}}}
\def\phantomvrule#1{\vbox{\hbox to0pt{\vrule width\tableaurule
height#1\hss}}}
\def\sqr{\vbox{%
  \phantomhrule\tableaustep
\hbox{\phantomvrule\tableaustep\kern\tableaustep\phantomvrule\tableaustep}%
  \hbox{\vbox{\phantomhrule\tableauside}\kern-\tableaurule}}}
\def\squares#1{\hbox{\count0=#1\noindent\loop\sqr
  \advance\count0 by-1 \ifnum\count0>0\repeat}}
\def\tableau#1{\vcenter{\offinterlineskip
  \tableaustep=\tableauside\advance\tableaustep by-\tableaurule
  \kern\normallineskip\hbox
    {\kern\normallineskip\vbox
      {\gettableau#1 0 }%
     \kern\normallineskip\kern\tableaurule}%
  \kern\normallineskip\kern\tableaurule}}
\def\gettableau#1 {\ifnum#1=0\let\next=\null\else
  \squares{#1}\let\next=\gettableau\fi\next}
\newcommand{\ex}{\mathrm{e}}
\newcommand{\esp}{\mathrm{e}}
\newcommand{\nordg}[1]{:\! #1 \!:_g} 
 
\newcommand{\nord}[1]{:\! #1 \!:} 
\newcommand{\gbullet}{\!\bullet\!}
\newcommand{\wbullet}{\!\circ\!}

\def\XXint#1#2#3{{\setbox0=\hbox{$#1{#2#3}{\int}$}
     \vcenter{\hbox{$#2#3$}}\kern-.5\wd0}}

\usepackage{tikz}
\usetikzlibrary{graphs}
\usetikzlibrary{trees}
\usetikzlibrary{calc}
\usetikzlibrary{decorations.markings}

\tikzstyle{gauge} = [circle, text centered, draw=black, minimum height=1.5cm]
\tikzstyle{flavor} = [rectangle, text centered, draw=black, minimum height=1.5cm,minimum width=1.5cm]

\tikzstyle{gaugeS} = [circle, text centered, draw=black, minimum height=6ex]
\tikzstyle{flavorS} = [rectangle, text centered, draw=black,minimum height=6ex,minimum width=6ex]
\usepackage{siunitx}

\makeatletter
\pgfdeclareshape{barn}{
    \inheritsavedanchors[from={rectangle}]
    \inheritanchorborder[from={rectangle}] 
    \inheritanchor[from={rectangle}]{center}
    \inheritanchor[from={rectangle}]{south}
    \inheritanchor[from={rectangle}]{west}
    \inheritanchor[from={rectangle}]{east}
    \inheritbackgroundpath[from={rectangle}]
   
    \backgroundpath{
    \southwest \pgf@xa=\pgf@x \pgf@ya=\pgf@y
    \northeast \pgf@xb=\pgf@x \pgf@yb=\pgf@y
    \pgf@xc=\pgf@xb \advance\pgf@xc by-\pgf@xa     
    \pgf@yc=\pgf@yb \advance\pgf@yc by-\pgf@ya    
    \advance\pgf@yb by-0.5\pgf@yc
    \pgfpathmoveto{\pgfpoint{\pgf@xa}{\pgf@ya}}
    \pgfpathlineto{\pgfpoint{\pgf@xa}{\pgf@yb}}
    \pgfpatharc{180}{0}{.5\pgf@xc}  
    \pgfpathlineto{\pgfpoint{\pgf@xb}{\pgf@ya}}
    \pgfpathclose
 }
}
\makeatother

\tikzstyle{gaugedflavor} = [barn,draw, text centered, minimum height=1.5cm,minimum width=1.5cm,draw=black]
\tikzstyle{gaugedflavorS} = [barn,draw, text centered, minimum width=6ex,minimum height=6ex,draw=black]

\pgfdeclareshape{cross}{
  \inheritsavedanchors[from={rectangle}]
    \inheritanchorborder[from={rectangle}] 
     \inheritanchor[from={rectangle}]{north}
    \inheritanchor[from={rectangle}]{center}
    \inheritanchor[from={rectangle}]{south}
    \inheritanchor[from={rectangle}]{west}
    \inheritanchor[from={rectangle}]{east}
\inheritbackgroundpath[from={rectangle}]
    \backgroundpath{
    \draw[solid] (.6ex,-.6ex) -- (-.6ex,.6ex);
    \draw[solid] (.6ex,.6ex) -- (-.6ex,-.6ex);
    }
}

\title{\boldmath Correlators between Wilson loop and chiral operators in $\mathcal{N}=2$ conformal gauge theories}
   
\author[a,b]{M.~Bill\`o,}
\affiliation[a]{Universit\`a di Torino, \\
Dipartimento di Fisica and Arnold-Regge Center\\}

\affiliation[b]{I.\,N.\,F.\,N. - sezione di Torino, \\
Via P. Giuria 1, I-10125 Torino, Italy\\}
\emailAdd{billo@to.infn.it}

\author[a,b]{F. Galvagno,}
\emailAdd{galvagno@to.infn.it}

\author[a,c]{P. Gregori,}
\affiliation[c]{Service de Physique Th\'eorique et Math\'ematique \\
Universit\'e Libre de Bruxelles and International Solvay Institutes,\\
Campus de la Plaine, CP 231, B-1050 Bruxelles, Belgium\\}
\emailAdd{pgregori@ulb.ac.be}

\author[d,b]{and A.~Lerda\,}
\affiliation[d]{Universit\`a del Piemonte Orientale,\\
Dipartimento di Scienze e Innovazione Tecnologica and Arnold-Regge Center\\
Viale T. Michel 11, I-15121 Alessandria, Italy\\}
\emailAdd{lerda@to.infn.it}

\abstract{We consider conformal $\cN=2$ super Yang-Mills theories with gauge group SU$(N)$ and
$N_f=2N$ fundamental hypermultiplets in presence of a circular 1/2-BPS Wilson loop. 
It is natural to conjecture that the matrix model which describes the expectation value of this system also encodes the one-point functions of chiral scalar operators in presence of the Wilson loop. We obtain 
evidence of this conjecture by successfully comparing, at finite $N$ and at the two-loop order, 
the one-point functions computed in field theory with the vacuum expectation values of 
the corresponding normal-ordered operators in the matrix model. For the part of these expressions 
with transcendentality $\zeta(3)$, we also obtain results in the large-$N$ limit that are exact in the 't Hooft coupling $\lambda$.
}


\preprint{ARC-18-03}

\begin{document}
\maketitle
\flushbottom

\section{Introduction}
\label{sec:intro}
The study of defects and of their properties may improve our understanding of quantum field theories. Wilson loops represent a class of gauge-invariant line defects which is of paramount relevance in gauge theories.

In general, Wilson loops receive perturbative and non-perturbative corrections and their exact evaluation 
is a difficult task. It is therefore important to find classes of theories and of Wilson loops for which it is 
possible to make progress in this direction. In this perspective, much work has been devoted to the study of Wilson loops in supersymmetric gauge theories, in theories which possess integrable sectors and in theories enjoying conformal symmetry. Furthermore, a powerful angle of approach to the strong coupling behavior is furnished by correspondences of the AdS/CFT type.
   
$\cN=4$ super Yang-Mills (SYM) theory is maximally supersymmetric, it is conformal and many sectors of its observables are integrable. Moreover, it is the theory for which the AdS/CFT correspondence was originally conjectured and for which it is best established. In this theory important results, many of which are exact, have been obtained regarding Wilson loop operators which preserve at least a fraction of the supersymmetry.  
In particular the 1/2-BPS circular Wilson loop has been evaluated exactly in terms of a Gaussian matrix 
model in \cite{Erickson:2000af,Drukker:2000rr,Pestun:2007rz}. 
Wilson loops preserving fewer supersymmetries \cite{Zarembo:2002an}, such as the 1/4-BPS circular loop \cite{Drukker:2006ga} and particular classes of 1/8-BPS loops 
\cite{Drukker:2007yx,Drukker:2007dw,Drukker:2007qr}, have been classified and analyzed. 
Correlators among such Wilson loops, or between Wilson loops and local operators have also been considered \cite{Semenoff:2001xp,Pestun:2002mr,Semenoff:2006am}; in particular, correlators of a 1/8-BPS circular loop and chiral primaries in $\mathcal{N}=4$ SYM theory have been computed \cite{Giombi:2009ds,Giombi:2012ep,Bassetto:2009rt,Bassetto:2009ms,Bonini:2014vta}, mapping them to multi-matrix models. Also correlators with local chiral operators and Wilson loops in higher representations have been discussed \cite{Giombi:2006de,Gomis:2008qa}.
Often these results have been successfully compared, at least in the large-$N$ limit, with AdS/CFT \cite{Berenstein:1998ij,Giombi:2006de,Gomis:2008qa} and with the outcome of the integrability approach \cite{Drukker:2012de}. 

$\cN=4$ SYM is a superconformal theory, and Wilson loops that preserve a subgroup of the superconformal symmetry are instances \cite{Kapustin:2005py} of a defect conformal field theory
(DCFT) \cite{McAvity:1993ue,McAvity:1995zd,Billo:2016cpy,Gadde:2016fbj}. 
The spectrum and the structure constants of operators defined on the defect represent an extra 
important piece of conformal data; correlators of certain such operators have been considered both 
directly \cite{Cooke:2017qgm,Kim:2017sju} and via integrability \cite{Cavaglia:2018lxi}. Also the 
correlators of the Wilson loop defect with bulk operators, such as the chiral primaries, are constrained by the residual symmetry. 

Similar progress has been made also in $\cN=2$ SYM theories, 
mainly thanks to localization techniques \cite{Teschner:2014oja,Pestun:2016zxk}.
These techniques, relying on supersymmetry, yield exact results for the field theory partition function in a deformed space-time geometry by localizing it on a finite set of critical points and expressing it as a matrix model. This procedure was extended by Pestun in a seminal paper \cite{Pestun:2007rz} to compute the expectation value of a circular Wilson loop in a $S^4$ sphere background, reducing the path integral computation to a matrix model which is a simple modification of the one for the partition function. 
In the $\mathcal{N}=4$ SYM case the matrix model is Gaussian, in agreement with the field theory 
results \cite{Drukker:2000rr,Erickson:2000af} mentioned above, while in the $\mathcal{N}=2$ theory it 
receives both one-loop and instanton corrections. 

Pestun's results on circular Wilson loops have opened several directions in the study of gauge theories
and allowed us to deepen our knowledge about the AdS/CFT duality in the $\mathcal{N}=2$ 
setting \cite{Rey:2010ry, Passerini:2011fe,Russo:2013sba}, as well as to provide exact results for some observables directly related to the Wilson loop, such as the Bremsstrahlung function \cite{Correa:2012at,Correa:2012hh,Lewkowycz:2013laa,Fiol:2015spa,Bonini:2015fng}.  

When the $\cN = 2$ theory is conformal, as it is the case for $\cN=2$ SQCD with 
$N_f = 2N$, it has been shown that the matrix model for the partition function on $S_4$ also contains information about correlators of chiral operators on $\mathbb{R}^4$ \cite{Baggio:2014ioa,Baggio:2014sna,Gerchkovitz:2014gta,Baggio:2015vxa,Baggio:2016skg}, provided one disentangles 
the operator mixing induced by the map from $S^4$ to $\mathbb{R}^4$ \cite{Gerchkovitz:2016gxx, Rodriguez-Gomez:2016ijh,Rodriguez-Gomez:2016cem}. 
In \cite{Billo:2017glv} this disentangling of operators has been realized as a
normal-ordering procedure and the relation between field theory and matrix model correlators has been
shown to hold also in non-conformal situations for a very special class of operators. 

It is natural to conjecture that, as it is the case in the $\cN=4$ theory, also in superconformal $\cN=2$ theories the matrix model for the circular Wilson loop on $S_4$ may contain information on correlators of chiral operators in the presence of a circular loop in $\mathbb{R}^4$.
In particular, from DCFT we know that the functional form of the one-point function in presence of a Wilson loop is completely fixed up to a coefficient depending on the coupling constant $g$; this coefficient can be encoded in the Pestun matrix model. 

In this paper, neglecting non-perturbative instanton contributions, we deal with the determinant factor in 
the matrix model definition, which can be expanded in powers of $g$. We work at finite and 
generic $N$. Following \cite{Billo:2017glv}, we identify the matrix model counterparts of chiral operators in the field theory through a normal-ordering prescription, and compute the one-point functions of such operators in the matrix model. We then compare them with the corresponding field theory one-point functions in presence of the Wilson loop computed in standard perturbation theory up to two loops for finite $N$, and to all orders in perturbation theory in planar limit for the $\zeta(3)$ dependent part. 
We heavily rely on the $\cN=4$ results in that we consider the diagrammatic difference between  
$\mathcal{N}=4$ and $\mathcal{N}=2$ \cite{Andree:2010na}; this procedure massively reduces the 
number of Feynman diagrams to be computed. We find complete agreement between the matrix model and field theory results; we believe that this represent compelling evidence for the conjecture. 

The paper is structured as follows. We introduce our set-up in Section~\ref{sec:WL}. In 
Sections~\ref{sec:mm} and \ref{secn:mmcwl} we perform the matrix model computation, reviewing 
first the $\cN=4$ case and then moving to  the superconformal $\cN=2$ theory. We also derive 
large-$N$ results which are exact in $\lambda= g N^2$ for the $\cN=4$ part of these one-point functions and for the extra part in the $\cN=2$ theory which has 
$\zeta(3)$ transcendentality.
The diagrammatic evaluation of the correlators in field theory is performed in Section~\ref{sec:pert}, 
up to two loops for finite $N$. We also show how the large-$N$ results derived in the matrix model approach arise diagrammatically. Finally, Section \ref{secn:concl} contains our conclusions, while some more technical
material is contained in three appendices.

\section{Wilson loop and its correlators with chiral operators}
\label{sec:WL}

We consider a $\mathcal{N}=2$ SYM theory on $\mathbb{R}^4$
with gauge group SU($N$) and $N_f$ fundamental flavours.
As is well-known, when $N_f=2N$ this theory is superconformal invariant, even at the quantum level. 
In the following we will restrict to this case.

We place a 1/2-BPS Wilson loop in a representation $\mathcal{R}$ along 
a circle $C$ of radius $R$ inside $\mathbb{R}^4$. Such operator, which we denote $W_{\mathcal{R}}(C)$, measures the holonomy of the gauge field and the adjoint scalars around $C$ and represents a (conformal) defect in the theory. The explicit expression of $W_{\mathcal{R}}(C)$ is
\begin{equation}
\label{defWR}
W_{\mathcal{R}}(C)=\frac{1}{N}\Tr_{\!\mathcal{R}}\, \mathcal{P}
\exp \left\{g \oint_C d\tau \Big[\ii \,A_{\mu}(x)\,\dot{x}^{\mu}(\tau)
+R\,\theta^{I}(\tau)\phi_I(x)\Big]\right\}
\end{equation}
with $I=1,2$.
Here $g$ is the gauge coupling constant, $A_{\mu}$ is the gauge field and $\phi_I$ are the two (real) 
scalar fields of the $\mathcal{N}=2$ vector multiplet, while $\mathcal{P}$ denotes the path-ordering and 
$\Tr_{\!\mathcal{R}}$ the trace in the representation $\mathcal{R}$ of SU($N$).
If we take $\theta^{I}(\tau)= \delta^{I1}$, which is the standard choice for the scalar
coupling, and introduce the chiral and anti-chiral combinations
\begin{equation}
\varphi=\frac{1}{\sqrt{2}}\big(\phi_1+\ii\,\phi_2\big)~,~~
\bar\varphi=\frac{1}{\sqrt{2}}\big(\phi_1-\ii\,\phi_2\big)~,
\end{equation}
the Wilson loop (\ref{defWR}) becomes
\begin{equation}
\label{WLdef2}
W_{\mathcal{R}}(C)=\frac{1}{N}\Tr_{\!\mathcal{R}}\, \mathcal{P}
\exp \left\{g \oint_C d\tau \Big[\ii \,A_{\mu}(x)\,\dot{x}^{\mu}(\tau)
+\frac{R}{\sqrt{2}}\big(\varphi(x) + \bar\varphi(x)\big)\Big]\right\}~.
\end{equation}
For definiteness, from now on we will take the representation $\mathcal{R}$ to be the fundamental 
of SU($N$) and denote the corresponding Wilson loop simply as $W(C)$. Furthermore, we will 
use the symbol ``$\,\tr$'' for the trace in the fundamental representation.

We are interested in computing the correlators between the Wilson loop and the chiral operators of the
SYM theory. The latter are labeled by a vector of integers $\vec{n}=(n_1,n_2,\cdots,n_\ell)$ and take 
a multi-trace expression of the form
\begin{equation}
O_{\vec{n}}(x)=\tr \varphi^{n_1}(x)\,\tr \varphi^{n_2}(x)\cdots\tr \varphi^{n_\ell}(x)~.
\label{On}
\end{equation}
In our model, these are protected chiral primary operators with a conformal dimension $n$ given by
\begin{equation}
n=\sum_{k=1}^\ell n_k~,
\label{defn}
\end{equation}
and obey chiral ring relations. Equivalently, by expanding $\varphi(x) = \varphi^b(x)\,T^b$, where $T^b$
are the generators of SU($N$) in the fundamental representation normalized in such a way that
\begin{equation}
\tr \big(T^bT^c\big)=\frac{1}{2}\,\delta^{bc}~,~~~\tr T^b=0 \quad\mbox{with}~~b,c=1,\cdots,N^2-1~,
\label{normT}
\end{equation}
we can write 
\begin{equation}
\label{OnR}
O_{\vec{n}}(x) = R_{\vec{n}}^{\,b_1\dots b_n}\, \varphi^{b_1}(x)\ldots \varphi^{b_n}(x)
\end{equation}
where $R_{\vec{n}}^{\,b_1\dots b_n}$ is a totally symmetric 
$n$-index tensor whose expression is encoded%
\footnote{Explicitly,
\begin{equation*}
R_{\vec{n}}^{\,b_1\dots b_n} 
= \tr \big(T^{(b_1}\cdots T^{b_{n_1}}\big)~
\tr \big(T^{b_{n_1+1}}\cdots T^{b_{n_1+n_2}}\big)\ldots
\tr \big(T^{b_{n_1 + \ldots + n_{\ell-1}+1}}\cdots T^{b_n)}\big)
\end{equation*}
where the indices are symmetrized with strength 1. \label{footnote:R}}
in (\ref{On}). 

\begin{figure}[ht]
\begin{center}
\begingroup%
  \makeatletter%
  \providecommand\color[2][]{%
    \errmessage{(Inkscape) Color is used for the text in Inkscape, but the package 'color.sty' is not loaded}%
    \renewcommand\color[2][]{}%
  }%
  \providecommand\transparent[1]{%
    \errmessage{(Inkscape) Transparency is used (non-zero) for the text in Inkscape, but the package 'transparent.sty' is not loaded}%
    \renewcommand\transparent[1]{}%
  }%
  \providecommand\rotatebox[2]{#2}%
  \ifx\svgwidth\undefined%
    \setlength{\unitlength}{240bp}%
    \ifx\svgscale\undefined%
      \relax%
    \else%
      \setlength{\unitlength}{\unitlength * \real{\svgscale}}%
    \fi%
  \else%
    \setlength{\unitlength}{\svgwidth}%
  \fi%
  \global\let\svgwidth\undefined%
  \global\let\svgscale\undefined%
  \makeatother%
  \begin{picture}(1,0.47275008)%
    \put(0,0){\includegraphics[width=\unitlength,page=1]{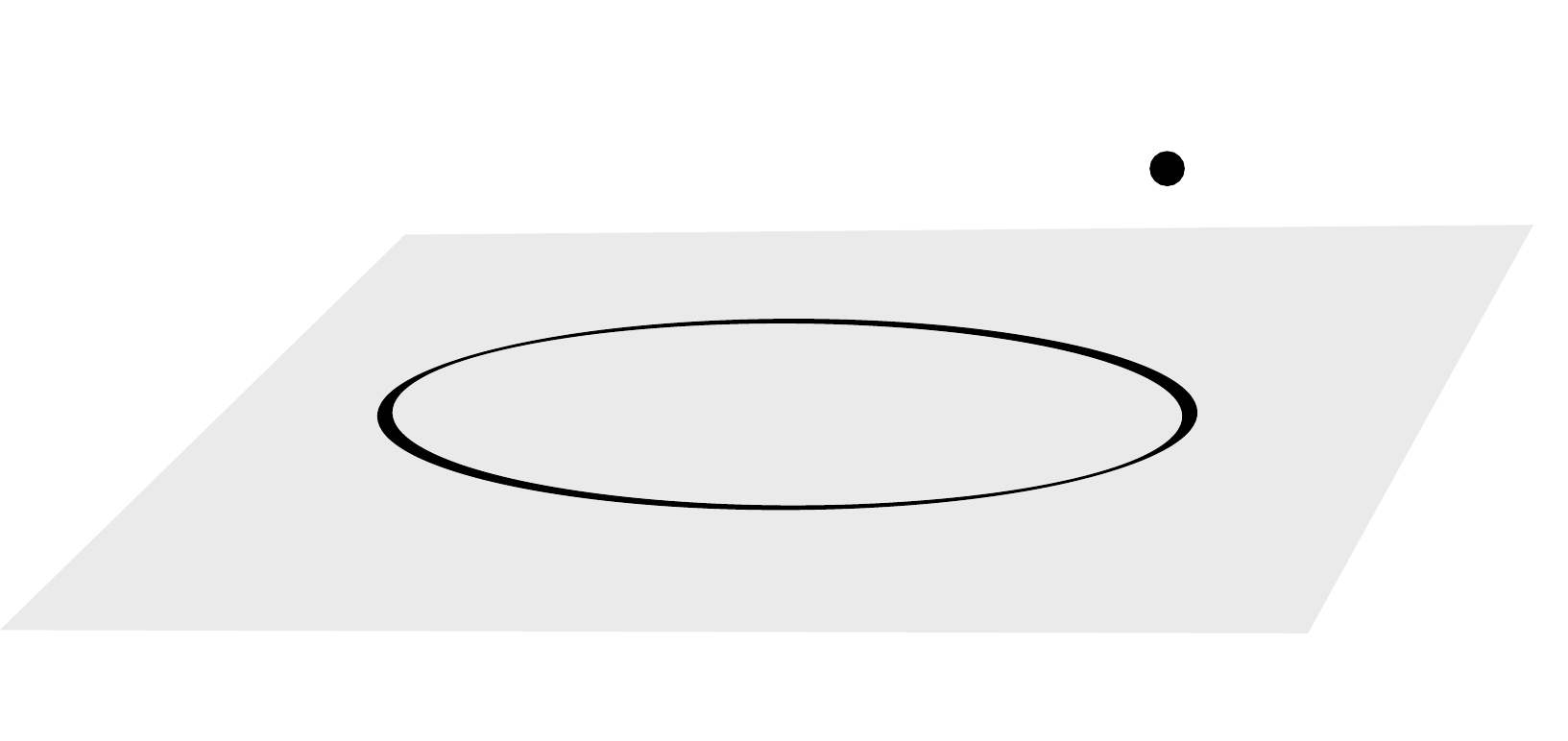}}%
    \put(0.7625012,0.37810715){\color[rgb]{0,0,0}\makebox(0,0)[lb]{\smash{$x$}}}%
    \put(0.1530294,0.12147429){\color[rgb]{0,0,0}\makebox(0,0)[lb]{\smash{$W(C)$}}}%
    \put(0,0){\includegraphics[width=\unitlength,page=2]{WOngeom.pdf}}%
    \put(0.92552756,0.16817663){\color[rgb]{0,0,0}\makebox(0,0)[lb]{\smash{$x_2$}}}%
    \put(0.34953029,0.0060181){\color[rgb]{0,0,0}\makebox(0,0)[lb]{\smash{$x_1$}}}%
    \put(0.50767365,0.45336258){\color[rgb]{0,0,0}\makebox(0,0)[lb]{\smash{$x_3,x_4$}}}%
    \put(0,0){\includegraphics[width=\unitlength,page=3]{WOngeom.pdf}}%
    \put(0.60292707,0.11029535){\color[rgb]{0,0,0}\makebox(0,0)[lb]{\smash{$r$}}}%
    \put(0.75362858,0.25823033){\color[rgb]{0,0,0}\makebox(0,0)[lb]{\smash{$L$}}}%
    \put(0,0){\includegraphics[width=\unitlength,page=4]{WOngeom.pdf}}%
    \put(0.35439669,0.21746727){\color[rgb]{0,0,0}\makebox(0,0)[lb]{\smash{$R$}}}%
  \end{picture}%
\endgroup%
\end{center}
\caption{The geometric set-up for the configuration we consider.}
\label{fig:WOngeom}
\end{figure}

The quantity of interest is the one-point function
\begin{equation}
\big\langle\, W(C)\,O_{\vec{n}}(x)\,\big\rangle~.
\label{WLO}
\end{equation}
To evaluate it, we can proceed as follows. Firstly, without any loss of generality, we can place the
circle $C$ in the plane $(x^1,x^2)\subset\mathbb{R}^4$. The points on the loop $C$ can then be parameterized as
\begin{equation}
x^\mu(\tau)=R\,\big(\cos\tau,\sin\tau,0,0\,\big)
\label{circle}
\end{equation}
with $\tau\in[\,0,2\pi\,]$. Secondly, using the standard results of defect conformal field theory 
\cite{Billo:2016cpy}, we can fix the functional dependence of the one-point function (\ref{WLO}). Indeed,
splitting the coordinates $x^\mu$ into parallel and transverse components, namely
$x^\mu\to(x^a;x^i)$ with $a=1,2$ and $i=3,4$, and denoting $x^ax_a=r^2$ and $x^ix_i=L^2$, so that
$x^2=r^2+L^2$ (see Fig.~\ref{fig:WOngeom}), we see that
\begin{equation}
\label{xCis}
\|x\|_C =\frac{\sqrt{\left(R^2 - x^2\right)^2 + 4 L^2 R^2}}{R}
\end{equation}
is the ``distance'' between $x$ and $C$, which is invariant under 
the $\mathrm{SO}(1,2)\times \mathrm{SO}(3)$ subgroup of the conformal symmetry that is preserved by the Wilson 
loop (see Appendix~\ref{secn:appa} for details).
When $x\to0$, we have $\|x\|_c\to R$. 

Because of conformal invariance, the correlator (\ref{WLO}) takes the form
\begin{equation}
\big\langle\,W(C)\,O_{\vec{n}}(x)\,\big\rangle= \frac{A_{\vec{n}}}{\,\big(2\pi\|x\|_C\big)^n\phantom{\Big|}}
\label{WLO1}
\end{equation}
where $A_{\vec{n}}$ is a $g$-dependent constant which corresponds 
to the one-point function evaluated in the origin:
\begin{equation}
A_{\vec{n}}= (2\pi R)^n\,\big\langle\, W(C)\,O_{\vec{n}}(0)\,\big\rangle~.
\label{Ang}
\end{equation}
In the next sections we will compute this function in two different ways: one by using the matrix model 
approach suggested by localization, and the other by using standard perturbative field theory methods. 
As anticipated in the Introduction, these two approaches lead to the same results.

\section{The matrix model approach}
\label{sec:mm}

The vacuum expectation value of the Wilson loop can be expressed and computed in terms 
of a matrix model, as shown in \cite{Pestun:2007rz} using localization methods. In the following we extend
this approach to compute also the correlators between the Wilson loop and the chiral correlators in
$\cN=2$ superconformal theories, but before we briefly review the matrix model and introduce our notations, 
relying mainly on \cite{Billo:2017glv}.

The matrix model in question corresponds to putting the $\cN=2$ 
SYM theory on a sphere $\cS_4$ and writing the
corresponding partition function as follows:
\begin{equation}
\label{ws444}
\cZ_{\mathcal{S}_4}=\int \prod_{u=1}^{N}\!da_u~\Delta(a)\,  \big| Z(\ii a)\big|^2 
\,\delta\Big(\sum_{v=1}^Na_v\Big)~. 
\end{equation}
Here $a_u$ are the eigenvalues of a traceless $N\times N$ matrix $a$ which are integrated over the real line; 
$\Delta(a)$ is the Vandermonde determinant and $Z(\ii a)$ is the gauge theory partition function 
on $\mathbb{R}^4$. The latter is computed using the localization techniques as in
\cite{Nekrasov:2002qd,Nekrasov:2003rj}, with the assumption that the adjoint scalar $\varphi(x)$
of the vector multiplet has a purely imaginary
vacuum expectation value given by $\langle\varphi\rangle=\ii\,a$, and that the $\Omega$-deformation parameters are $\epsilon_1=\epsilon_2=1/R$ where $R$ is the radius of $\cS_4$ which from now on
we take to be 1 for simplicity. This partition function is a 
product of the classical, 1-loop and instanton contributions, namely:
\begin{equation}
  Z(\ii a)  =  Z_{\mathrm{class}}(\ii a)\,  Z_{\mathrm{1-loop}} (\ii a)\,  Z_{\mathrm{inst}} (\ii a) ~.
\end{equation}
The classical part provides a Gaussian term in the matrix model:
 \begin{equation}
 \big|Z_{\mathrm{class}}(\ii a)\big|^2  
 = \, \ex^{-\frac{8\pi^2}{g^2} \,\tr a^2  } ~,
 \end{equation}
while the 1-loop contribution is
\begin{equation}
\big|Z_{\mathrm{1-loop}}(\ii a)\big|^2 \,=\, 
\prod_{u < v=1}^N \!\! H( \ii a_{uv} )^2\,\,\prod_{u=1 }^N  H( \ii a_{u} )^{-N_f}
\label{zloops4}
\end{equation}
where $a_{uv}=a_u-a_v$, and
\begin{equation}
H(x)= G(1+x)\, G(1-x)
\label{H}
\end{equation}
with $G(x)$ being the Barnes $G$-function.
In the weak-coupling limit $g \ll 1$, where instantons are exponentially suppressed, we can set
\begin{equation}
  \big|Z_{\mathrm{inst}}(\ii a)\big|^2  =1~. 
\end{equation}
Moreover, in this limit the integral (\ref{ws444}) is dominated by the region of small $a_u$, 
and thus we can expand the functions $H$ appearing in (\ref{zloops4}) using 
\begin{equation}
\log  H(x) =-(1+\gamma)\,x^2 -\sum_{n=2}^\infty \zeta(2n-1) \,\frac{x^{2n}}{n}
\label{hsmall}
\end{equation}
where $\zeta(n)$ is the Riemann zeta-function and $\gamma$ is the Euler-Mascheroni constant. 
In this way the one-loop contribution can be viewed as an interaction term in a free matrix model:
\begin{equation}
\big|Z_{\mathrm{1-loop}}(\ii a)\big|^2 = e^{-S_{\mathrm{int}}(a) }
\end{equation}
where $S_{\mathrm{int}}(a)$ is a sum of homogeneous polynomials $S_n$ in $a$ of order $n$.  
The first few are:
 \begin{equation}
 \begin{aligned}
S_2(a) &= -(1+\gamma) \,(2N-N_f)\, \tr a^2 =0~,\phantom{\frac{1}{2}}\\
S_4(a) &=  \frac{\zeta(3)}{2} \, \Big[ (2 N-N_f) \,
{\tr}\,a^4 + 6 \left(\tr a^2\right)^2  \Big] =3\,\zeta(3)\left(\tr a^2\right)^2 ~,  \\
S_6(a) &= - \frac{\zeta(5)}{3} \,  \Big[ (2 N-N_f) \, \tr a^6 +30\, \tr a^4 \, \tr a^2 -20 \,\left(\tr \,a^3\right)^2
 \Big]\\
 &= - \frac{10\,\zeta(5)}{3} \,  \Big[3\, \tr a^4 \, \tr a^2 -2\,\left(\tr a^3\right)^2
 \Big]
 \end{aligned} 
 \label{Sintexp}
 \end{equation}
 where the last step in each line follow from the superconformal condition $N_f=2N$.
After the rescaling 
\begin{equation}
\label{resca}
a \to \Big(\frac{g^2}{8\pi^2}\Big)^{\!\frac 12}\, a~,
\end{equation} 
the matrix model gets a canonically normalized Gaussian factor and the sphere partition function becomes
\begin{equation}
\label{rescaledmm}
\cZ_{\mathcal{S}_4} = \Big(\frac{g^2}{8\pi^2}\Big)^{\frac{N^2-1}{2}} \, \int \prod_{u=1}^{N} 
da_u~\Delta(a)\,\ex^{-\tr a^2 - S_{\mathrm{int}}(a)}\,\delta\Big(\sum_{v=1}^Na_v\Big)
\end{equation} 
with
\begin{equation}
\label{Sintresc}
S_{\mathrm{int}}(a) =\frac{3\,\zeta(3)\,g^4}{(8\pi^2)^2} \,\left(\tr a^2\right)^2  -
\frac{10\,\zeta(5)\,g^6}{3(8\pi^2)^3}\, 
\Big[3\, \tr a^4 \, \tr a^2 -2\,\left(\tr a^3\right)^2 \Big]+ \cdots ~.
\end{equation}
Exploiting the Vandermonde determinant $\Delta(a)$ and writing $a=a^b\,T^b$, we can alternatively express 
the integral (\ref{rescaledmm}) using a flat integration measure $da$ over all matrix components 
$a^b$ as follows
\begin{equation}
\label{Zmmgauss}
\cZ_{\mathcal{S}_4}\, = c_N \, \Big(\frac{g^2}{8\pi^2}\Big)^{\frac{N^2-1}{2}} \,  \int da~
\ex^{-\tr a^2 - S_{\mathrm{int}}(a)}  
\end{equation}
where $c_N$ is a $g$-independent constant and $da\propto \prod_{b}da_b$. The overall factor $c_N$ 
and the normalization of the flat measure $da$ are clearly irrelevant for the computation of the vacuum
expectation value of any quantity $f(a)$, which is defined as 
\begin{equation}
\label{vevmat}
\begin{aligned}
\big\langle\,f(a)\,\big\rangle\, & = \,\frac{\displaystyle{ \int \!da ~\esp^{-\tr a^2-S_{\mathrm{int}}(a)}\,
f(a)}}
{ \displaystyle{\int \!da~\esp^{-\tr a^2-S_{\mathrm{int}}(a)}} }
= \frac{\big\langle\,
\esp^{- S_{\mathrm{int}}(a)}\,f(a)\,\big\rangle_0\phantom{\Big|}}
{\big\langle\,\esp^{- S_{\mathrm{int}}(a)}\,\big\rangle_0
\phantom{\Big|}}~.
\end{aligned}
\end{equation} 
Here we have denoted with a subscript 0 the expectation value in the 
Gaussian matrix model, namely
\begin{equation}
\big\langle\,f(a)\,\big\rangle_0 =
\frac{\displaystyle{\int \!da\,\,\esp^{-\tr a^2} \, f(a)}}{\displaystyle{\int \!da\,\,\esp^{-\tr a^2}}}~.
\label{vevO}
\end{equation}
This Gaussian model is the matrix model that is appropriate to describe the $\cN=4$ SYM theory. 
In this case, in fact, the field content of the theory is such that 
the 1-loop partition function $Z_{\mathrm{1-loop}}$ and the instanton partition function 
$Z_{\mathrm{inst}}$ are both equal to 1, implying that $S_{\mathrm{int}}=0$.

Notice that if we normalize the flat measure as
\begin{equation}
\label{decadja}
 da = \prod_{b=1}^{N^2-1} \frac{da^b}{\sqrt{2\pi}}~,
\end{equation}
then the denominator of (\ref{vevO}) becomes 1 and we simply have
\begin{equation}
\big\langle\,f(a)\,\big\rangle_0 =\int \!da\,\,\esp^{-\tr a^2} \, f(a)~.
\label{vevO1}
\end{equation}
Using this, we can easily see that the basic Wick contraction in the Gaussian model is
\begin{equation}
\label{basicWicka}
\big\langle\, a^b \, a^c\,\big\rangle_0 = \delta^{bc} ~.
\end{equation} 
Introducing the notation
\begin{equation}
t_{n_1,n_2,\cdots} = \big\langle\,\tr a^{n_1}\,\tr a^{n_2} \cdots \big\rangle_0
\label{amptree}
\end{equation}
and using (\ref{normT}), we evidently have 
\begin{equation}
\label{rrrecursion-2}
\begin{aligned}
t_0 =\big\langle\, \tr 1\,\big\rangle_0 =N~,~~~
t_1 = \big\langle\, \tr a\,\big\rangle_0=0~,~~~
t_2 =  \big\langle\, \tr a^2\,\big\rangle_0 =\frac{N^2-1}{2}~.
\end{aligned}
\end{equation}
Higher order traces can be computed performing consecutive Wick contractions with (\ref{basicWicka})
and using the fusion/fission identities
\begin{equation}
\begin{aligned}
\tr \big(T^b B \,T^b C\big)&=\frac{1}{2}\,\tr B~\tr C-\frac{1}{2N}\,\tr \big(B\, C\big)~,\\
\tr \big(T^b C\big)~\tr \big(T^b C\big)&=\frac{1}{2}\,\tr \big(B\, C\big)-\frac{1}{2N}\,\tr B~\tr C~,
\end{aligned}
\label{fusionfission}
\end{equation}
which hold for any two matrices $B$ and $C$.
In this way we can build recursion relations and, for example, get:
\begin{align}
 t_{n}    &=   \frac12 \sum_{m=0}^{n-2}  \Big( t_{m,n-m-2}
  -\frac{1}{N}\, t_{n-2}  \Big)  ~,\notag\\
 t_{n,n_1} &=  \frac12 \sum_{m=0}^{n-2}  \Big( t_{m,n-m-2,n_1}
  -\frac{1}{N}\,   t_{n-2,n_1}  \Big)  
  + \frac{n_1}{2} \,\Big(  t_{n+n_1-2 } -\frac{1}{N} \,t_{n-1,n_1-1} \Big)~,\label{rrecursion}\\
  t_{n,n_1,n_2}  &=\frac12 \sum_{m=0}^{n-2}  \Big( t_{m,n-m-2,n_1,n_2}
  -\frac{1}{N} \,  t_{n-2,n_1,n_2}  \Big)  + \frac{n_1}{2} \Big(t_{n+n_1-2,n_2}
  -\frac{1}{N} \,  t_{n-1,n_1-1,n_2}\Big)  \notag\\
   &~~~~~~~~~~~+ \frac{n_2}{2} \Big( t_{n+n_2-2,n_1 }
   -\frac{1}{N}  t_{n-1,n_1,n_2-1} \Big) ~,\notag
\end{align} 
and so on. 
These  relations, together with the initial conditions (\ref{rrrecursion-2}), give an efficient way to obtain
multi-trace vacuum expectation values in the Gaussian model and will be the basic ingredients for the 
computations of the correlators in the $\cN=2$ superconformal theory.

\subsection{Wilson loop and chiral operators in the matrix model}
As shown in \cite{Pestun:2007rz}, in the matrix model the Wilson loop (\ref{WLdef2})
in the fundamental representation and on a circle of radius $R=1$ is given by the following operator
\begin{equation}
\cW(a)=\frac{1}{N}\,\tr\!\exp\Big(\frac{g}{\sqrt{2}}\,a\Big)=\frac{1}{N}
\sum_{k=0}^\infty \frac{g^k}{2^{\frac{k}{2}}\,k!}\,\tr a^k ~.
\label{WLmm}
\end{equation}
On the other hand, to any multi-trace chiral operator $O_{\vec{n}}(x)$ of the SYM theory defined 
as in (\ref{On}), it would seem natural to associate a matrix operator 
$O_{\vec{n}}(a)$ with precisely the same expression but with the field $\varphi(x)$ replaced by 
the matrix $a$, namely
\begin{equation}
O_{\vec{n}}(a)=\tr a^{n_1}\,\tr a^{n_2}\cdots\tr a^{n_\ell}\,=\,R_{\vec{n}}^{\,b_1\dots b_n}\,
\,a^{b_1}\,a^{b_2}\cdots a^{b_n}~.
\label{Ona}
\end{equation}
However, since the field theory propagator only connects $\varphi$ with $\bar\varphi$, 
all operators $O_{\vec{n}}(x)$ have no self-contractions, whereas the operators 
$O_{\vec{n}}(a)$ defined above do not share this property. This means that the dictionary
between the SYM theory and the matrix model is more subtle. Indeed, we have to subtract from
$O_{\vec{n}}(a)$ all its self-contractions by making it orthogonal to all the lower dimensional operators, or
equivalently by making it normal-ordered. As discussed in \cite{Billo:2017glv}, given any operator 
$O(a)$ we can define its normal-ordered version $\cO(a)$ as follows. Let be $\Delta$ the dimension of
$O(a)$ and $\big\{O_p(a)\big\}$ a basis of in the finite-dimensional space of matrix operators with dimension smaller than $\Delta$. Denoting by $C_\Delta$ the (finite-dimensional) matrix of correlators
\begin{equation}
\big(C_\Delta\big)_{pq} = \big\langle\,O_p(a)\,O_q(a)\,\big\rangle
\label{Cpq}
\end{equation}
which are computed according to (\ref{vevmat}), we define the normal-ordered operator 
\begin{equation}
\cO(a) =\,\, \nordg{O(a)}\,\,
=\, O(a) - \sum_{p,q} \big\langle\,O(a)\, O_{p}(a)\,\big\rangle\, (C_\Delta^{-1})^{pq}\, O_q(a)~.
\label{normalo}
\end{equation}
As emphasized by the notation, the normal-ordered operators are $g$-dependent, since the 
correlators in the right hand side of (\ref{normalo}) are computed in the interacting $\cN=2$ matrix model
using (\ref{Sintresc}).

Using these definitions, the correspondence between field theory and matrix model operators takes
the following simple form
\begin{equation}
\label{corrOcO}
O_{\vec{n}}(x) \,\to\, \cO_{\vec{n}}(a) =\,\, \nordg{O_{\vec{n}}(a)}~.
\end{equation} 
Let us give some explicit examples by considering the first few low-dimensional operators. At level $n=2$
we have just one operator:
\begin{equation}
\label{noO2} 
\cO_{(2)}(a) = \,\,\nordg{\tr a^2}\,\,= \tr a^2-\frac{N^2-1}{2}
+\frac{3\,\zeta(3)\,g^4}{(8\pi^2)^2}\,\frac{(N^2-1)(N^2+1)}{2}+O(g^6)~.
\end{equation} 
Similarly, at level $n=3$ we have one operator, which in the SU($N$) theory does not receive any correction:
\begin{equation}
\label{noO3}
\cO_{(3)}(a) = \,\,\nordg{\tr a^3}\,\,= \tr a^3~.
\end{equation}
At level $n=4$, we have instead two independent operators corresponding to 
$\vec{n}=(4)$ and $\vec{n}=(2,2)$. Their normal-ordered expressions are given, respectively, by
\begin{align}
\cO_{(4)}(a) &= \,\,\nordg{\tr a^4}\nonumber\\
&= \tr a^4 
-\frac{2N^2-3}{N}\,\tr a^2+\frac{(N^2-1)(2N^2-3)}{4N}
\phantom{\Big|}\label{noO4}\\
&~~+\frac{3\,\zeta(3)\,g^4}{(8\pi^2)^2}
\Big[\frac{(2N^2-3)(N^2+5)}{N}\:\tr a^2-\frac{2(N^2-1)(N^2+4)(2N^2-3)}{4N}\Big] 
+O(g^6)~,\phantom{\Big|}\nonumber
\end{align}
and
\begin{align}
\cO_{(2,2)}(a) &= \,\,\nordg{\left(\tr a^2\right)^2}\nonumber \\
&=\left(\tr a^2\right)^2-(N^2-1)\,\tr a^2+\frac{N^4-1}{4}
\phantom{\Big|}\label{noO22}\\
&~~+\frac{3\,\zeta(3)\,g^4}{(8\pi^2))^2}\Big[
(N^2-1)(N^2+5)\,\tr a^2-\frac{(N^4-1)(N^2+4)}{2}\Big]
+\mathcal{O}(g^6)~.\phantom{\Big|}\nonumber
\end{align}
 Up to the order $g^6$ we have considered, it is easy to check that these operators satisfy
\begin{equation}
\begin{aligned}
\big\langle\,\cO_{\vec{n}}(a)\,\big\rangle&=0~,\\
\big\langle\,\cO_{\vec{n}}(a)\,\cO_{\vec{m}}(a)\,\big\rangle&=0~,\\
\end{aligned}
\end{equation}
for $n\neq m$. 
Normal-ordered operators of higher dimension can be constructed without any problem along these same
lines.

We observe that the $g$-independent parts of the above expressions correspond to the normal-ordered 
operators in the Gaussian model, {\it{i.e.}} in the $\cN=4$ theory. Since we will often compare our
$\cN=2$ results with those of the $\cN=4$ theory, we find convenient to introduce a specific notation
for the $g\to 0$ limit of the normal ordering and write
\begin{equation}
\widehat{\cO}_{\vec{n}}(a) \equiv \lim_{g\to 0} \cO_{\vec{n}}(a) =\,\,\, \nord{O_{\vec{n}}(a)}~,
\label{hatOn}
\end{equation}
so that most of the formulas will look simpler.

In the following section we will explicitly compute the one-point functions between the Wilson loop and
the chiral operators in the $\cN=2$ matrix model, namely
\begin{equation}
\cA_{\vec{n}}=\big\langle\,\cW(a)\,\cO_{\vec{n}}(a)\,\big\rangle
\label{Angdef}
\end{equation}
which will later compare with the field theory amplitudes defined in (\ref{Ang}).

\section{Matrix model correlators in presence of a Wilson loop}
\label{secn:mmcwl}

Our main goal here is the computation of $\cA_{\vec{n}}$ 
in the interacting matrix model described above. 
As a warming-up, but also for later applications, we begin by presenting the results in the 
Gaussian matrix model, {\it{i.e.}} in the $\cN=4$ theory.

\subsection{The $\cN=4$ theory}
\label{subsubsec:N4mmc}
In this case we should consider the operators $\widehat{\cO}_{\vec{n}}(a)$ defined in (\ref{hatOn}) and
compute
\begin{equation}
\widehat{\cA}_{\vec{n}}=\big\langle\,\cW(a)\,\widehat{\cO}_{\vec{n}}(a)\,\big\rangle_0
\label{hatAn}
\end{equation}
using the definition (\ref{vevO1}).

The simplest example is the amplitude with the identity ($\vec{n}=(0)$), 
which yields the vacuum expectation value of the Wilson loop operator (\ref{WLmm}):
\begin{equation}
\label{WvevN4}
\widehat{\cA}_{(0)}=
\big\langle\,\cW(a)\,\big\rangle_0=\frac{1}{N}\,\sum_{k=0}^\infty
\frac{g^k}{2^{\frac{k}{2}}\,k!}\,t_k 
\end{equation}
with $t_k$ defined in (\ref{amptree}). Using the explicit expressions given in (\ref{rrrecursion-2}) 
and (\ref{rrecursion}), we find
\begin{equation}
\begin{aligned}
\!\!\!\widehat{\cA}_{(0)}&\!=1+g^2\,\frac{N^2-1}{8N}+g^4\,\frac{(N^2-1)(2N^2-3)}{384N^2}
+g^6\,\frac{(N^2-1)(N^4-3N^2+3)}{9216N^3}+\cdots
\end{aligned}
\label{hatA0exp}
\end{equation}
This perturbative series can be resummed into
\begin{align}
\widehat{\cA}_{(0)}
=\frac{1}{N}\,L_{N-1}^{1}\Big(-\frac{g^2}{4}\Big)\,\exp\Big[\frac{g^2}{8}\Big(1-\frac{1}{N}\Big)\Big]
\label{WvevN4a}
\end{align}
where $L_n^m$ is the generalized Laguerre polynomial of degree $n$. This is the SU($N$) version of the
well-known result of \cite{Drukker:2000rr}, originally derived for U($N$).

Next we consider the amplitude between the Wilson loop and the operator $\widehat{\cO}_{(2)}(a)$ at level
2. This is given by
\begin{equation}
\widehat{\cA}_{(2)}=\big\langle\,\cW(a)\,\nord{\tr a^2}\big\rangle_0
=\frac{1}{N}\,\sum_{k=0}^\infty
\frac{g^k}{2^{\frac{k}{2}}\,k!}\,\Big(t_{k,2}-\frac{N^2-1}{2}\,t_k\Big) ~.
\label{A20}
\end{equation}
The recursion relations (\ref{rrecursion}) imply 
\begin{equation}
t_{k,2}=\Big(\frac{k}{2}+\frac{N^2-1}{2}\Big)t_k~,
\end{equation}
and thus the amplitude (\ref{A20}) becomes
\begin{equation}
\widehat{\cA}_{(2)}=\frac{1}{N}\,\sum_{k=0}^\infty\frac{k}{2}\,
\frac{g^k}{2^{\frac{k}{2}}\,k!}\,t_k =\frac{g}{2}\,\partial_g\widehat{\cA}_{(0)}~.
\label{A21}
\end{equation}
Expanding for small $g$, we get
\begin{equation}
\widehat{\cA}_{(2)}=g^2\,\frac{N^2-1}{8N}+g^4\,\frac{(N^2-1)(2N^2-3)}{192N^2}
+g^6\,\frac{(N^2-1)(N^4-3N^2+3)}{3072N^3}+\cdots~.
\label{A21g}
\end{equation}
This same procedure can be used to compute the amplitudes $\widehat{\cA}_{\vec{n}}$ for
any $\vec{n}$. The remarkable fact is that, thanks to the recursion relations (\ref{rrecursion}),
it is always possible to obtain compact expressions in terms of $\widehat{\cA}_{(0)}$ and its
derivatives that are exact, {\it{i.e.}} valid for any $N$ and any $g$. For example, at level $n=3$ we find
\begin{equation}
\widehat{\cA}_{(3)}=\frac{g}{\sqrt{2}}\,\partial_g^2\widehat{\cA}_{(0)}-\frac{g^2}{4\sqrt{2}N}
\,\partial_g\widehat{\cA}_{(0)}-\frac{g(N^2-1)}{4\sqrt{2}N}
\,\widehat{\cA}_{(0)}~,
\label{A3}
\end{equation}
while at level $n=4$ we have
\begin{equation}
\widehat{\cA}_{(4)}=g\,\partial_g^3\widehat{\cA}_{(0)}+\frac{g^2}{4N}\,\partial_g^2\widehat{\cA}_{(0)}
+\frac{g^3-4gN(2N^2-3)}{16N^2}\,\partial_g\widehat{\cA}_{(0)}+
\frac{g^2(N^2-1)}{16N^2}\,\widehat{\cA}_{(0)}~,
\label{A4}
\end{equation}
and
\begin{equation}
\widehat{\cA}_{(2,2)}=\frac{g^2}{4}\,\partial_g^2\widehat{\cA}_{(0)}
-\frac{g}{4}\,\partial_g\widehat{\cA}_{(0)}~.
\end{equation}
We have performed similar calculations for higher dimensional operators, but we do not report the
results since they would not add much to what we have already exhibited. Instead, we point out that
the lowest order term in the small $g$ expansion of $\widehat{\cA}_{\vec{n}}$, 
which we call ``tree-level term'', can be compactly written as
\begin{equation}
\begin{aligned}
\widehat{\cA}_{\vec{n}}\Big|_{\mathrm{tree-level}}\,&=\frac{g^n}{N\,2^{\frac{n}{2}}\,n!}\,
R_{\vec{n}}^{\,b_1\dots b_n}\,
\big\langle\,\tr a^n\,\nord{a^{b_1}\dots \,a^{b_n}}\big\rangle_0\\
&=\frac{g^n}{N\,2^{\frac{n}{2}}}\,R_{\vec{n}}^{\,b_1\dots b_n}\,\tr \big(T^{b_1}\dots T^{b_n}\big)
\end{aligned}
\label{Antree}
\end{equation}
where $R_{\vec{n}}^{\,b_1\dots b_n}$ is the symmetric tensor 
associated to the operator $O_{\vec{n}}(a)$ according to (\ref{Ona}).
For later convenience, in Tab.~\ref{tab1} we collect the explicit expressions of 
$\widehat{\cA}_{\vec{n}}\big|_{\mathrm{tree-level}}$ for all operators up to level $n=4$.
\begin{table}[ht]
\begin{center}
{
\begin{tabular}{|c|c|}
\hline
$\vec{n}\phantom{\bigg|}$&$\widehat{\cA}_{\vec{n}}\big|_{\mathrm{tree-level}}$\\
      \hline\hline
$(2)\phantom{\bigg|}$ & $g^2\frac{N^2-1}{8N\phantom{\big|}}$ \\
\hline
$(3)\phantom{\bigg|}$ & $g^3\frac{(N^2-1)(N^2-4)}{32\sqrt{2}N^2\phantom{\big|}}$ \\
\hline
$(4)\phantom{\bigg|}$ & $g^4\frac{(N^2-1)(N^4-6N^2+18)}{384 N^3\phantom{\big|}}$ \\
\hline
$(2,2)\phantom{\bigg|}$ & $g^4\frac{(N^2-1)(2N^2-3)}{192N^2\phantom{\big|}}$ \\
\hline
\end{tabular}
}
\end{center}
\caption{The tree-level contribution to $\widehat{\cA}_{\vec{n}}$ for operators up to order $n=4$.}
\label{tab1}
\end{table}

\subsection{The $\mathcal{N}=2$ superconformal theory}

Let us now return to our main goal, namely the computation of the one-point amplitudes in 
the interacting matrix model that describes the $\cN=2$ superconformal theory.
Comparing $\cA_{\vec{n}}$ with the $\cN=4$ amplitudes $\widehat{\cA}_{\vec{n}}$, we see
two main differences:
\begin{enumerate}
\item the normal-ordered operators $\cO_{\vec{n}}$ explicitly contain $g$-dependent terms;
\item the vacuum expectation value is computed in a $g$-dependent matrix model.
\end{enumerate}
Both effects arise from the interaction terms of $S_{\mathrm{int}}(a)$ given in (\ref{Sintresc}); 
thus we can write
\begin{equation}
\cA_{\vec{n}}=\widehat{\cA}_{\vec{n}}+\delta\cA_{\vec{n}}
\label{diffAn}
\end{equation}
with
\begin{equation}
\delta\cA_{\vec{n}}=\frac{3\,\zeta(3)\,g^4}{(8\pi^2)^2}\,\cX_{\vec{n}}
-\frac{10\,\zeta(5)\,g^6}{3(8\pi^2)^3}\,\cY_{\vec{n}}+\,\cdots
\label{deltaAnexp}
\end{equation}
where the ellipses stand for terms of higher transcendentality,
proportional to  $\zeta(7)$, $\zeta(3)^2$ and so on. 
The quantities $\cX_{\vec{n}}$, $\cY_{\vec{n}}$ and the analogous ones at higher transcendentality
depend on the coupling constant $g$ and can be expressed using vacuum expectation values in the 
Gaussian model and, eventually, $\widehat{\cA}_{(0)}$ and its derivatives in a compact way.
Since $\delta\cA_{\vec{n}}$ starts at order $g^4$, {\it{i.e.}} at two loops, we clearly have
\begin{equation}
\delta\cA_{\vec{n}}\Big|_{\mathrm{tree-level}}=0\qquad\mbox{and}\qquad
\delta\cA_{\vec{n}}\Big|_{\mathrm{1-loop}}=0\label{deltaAntree1loop}
\end{equation}
for any $\vec{n}$.
In the following we will restrict our analysis to the first correction $\cX_{\vec{n}}$ for which we will 
provide explicit formulas in several examples.

Let us start with the Wilson loop, {\it{i.e.}} with the identity operator ($n=0$). In this case there 
is no normal-ordering to do and thus the only contribution to $\cX_{(0)}$ comes from the
interactions in the matrix model. Focusing on the $\zeta(3)$-term which is proportional 
to $\left(\tr a^2\right)^2$, after some straightforward algebra we get
\begin{equation}
\begin{aligned}
\cX_{(0)}&=-\,\big\langle\, \cW(a)\left(\tr a^2\right)^2\big\rangle_0+
\big\langle\, \cW(a)\,\big\rangle_0~\big\langle\, \left(\tr a^2\right)^2\big\rangle_0~.
\end{aligned}
\label{X0}
\end{equation}
Evaluating the vacuum expectation values by means of 
the recursion relations (\ref{rrecursion}) and expressing the results in terms of the $\cN=4$ Wilson loop, 
we can rewrite the above expression as
\begin{equation}
\cX_{(0)}=-\frac{g^2}{4}\,\partial_g^2\widehat{\cA}_{(0)}-\frac{g(2N^2+1)}{4}\,\partial_g\widehat{\cA}_{(0)}~.
\end{equation}
Using (\ref{WvevN4a}) and expanding for small $g$, we easily get
\begin{equation}
\begin{aligned}
\cX_{(0)}&=-\,g^2\,\frac{(N^2-1)(N^2+1)}{8N}-g^4\,\frac{(N^2-1)(2N^2-3)(N^2+2)}{192N^2}\\
&~~~\,-g^6\,\frac{(N^2-1)(N^4-3N^2+3)(N^2+3)}{8N}
+\cdots~.
\end{aligned}
\end{equation}
Therefore, in the difference $\delta\cA_{(0)}$ the leading term, which is a 2-loop effect
induced by the $g^4$-part of $S_{\mathrm{int}}(a)$ proportional to $\zeta(3)$, 
turns out to be
\begin{equation}
\delta\cA_{(0)}\Big|_{\mathrm{2-loop}}= -g^6\,\frac{\zeta(3)}{(8\pi^2)^2}\,
\frac{3(N^2-1)(N^2+1)}{8N}~.
\label{deltaA02loop}
\end{equation}
This expression has been successfully checked in \cite{Andree:2010na} against an explicit perturbative
2-loop calculation in field theory.

Let us now consider the operator $\cO_{(2)}$ at level $n=2$. In this case we have
\begin{align}
\cX_{(2)}&=
-\,\big\langle\, \cW(a)\,\widehat{\cO}_{(2)}(a)\left(\tr a^2\right)^2\big\rangle_0+
\big\langle\, \cW(a)\,\widehat{\cO}_{(2)}(a)\,\big\rangle_0~\big\langle \!
\left(\tr a^2\right)^2\big\rangle_0\notag\phantom{\Big|}\\
&~~~\,+\frac{(N^2-1)(N^2+1)}{2}\,\big\langle\,\cW(a)\,\big\rangle_0
\label{X2}
\end{align}
where the last term is due to the normal-ordering procedure in the interacting theory which indeed
yields a part proportional to $(N^2-1)(N^2+1)/2$ (see (\ref{noO2})).
Evaluating the vacuum expectation values, this expression becomes
\begin{equation}
\cX_{(2)}=
-\frac{g^3}{8}\,\partial_g^3\widehat{\cA}_{(0)}
-\frac{g^2(2N^2+7)}{8}\,\partial_g^2\widehat{\cA}_{(0)}
-\frac{5g(2N^2+1)}{8}\,\partial_g\widehat{\cA}_{(0)}~,
\label{X2a}
\end{equation}
while its perturbative expansion is
\begin{equation}
\begin{aligned}
\cX_{(2)}&=-g^2\,\frac{3(N^2-1)(N^2+1)}{8N}-g^4\,\frac{(N^2-1)(2N^2-3)(N^2+2)}{48N^2}\\
&~~~\,-g^6 \,\frac{5(N^2-1)(N^4-3 N^2+3)(N^2+3)}{3072 N^3}+\cdots~.
\end{aligned}
\label{X2b}
\end{equation}
The leading term tells us that the 2-loop correction to the $\cN=2$ amplitude $\cA_{(2)}$ is
\begin{equation}
\delta\cA_{(2)}\Big|_{\mathrm{2-loop}}= -g^6\,\frac{\zeta(3)}{(8\pi^2)^2}\,
\frac{9(N^2-1)(N^2+1)}{8N}~.
\label{deltaA22loop}
\end{equation}

This procedure can be easily applied to operators of higher dimensions. 
For example, skipping the intermediate steps, at level $n=3$
we find
\begin{equation}
\begin{aligned}
\cX_{(3)}
&=-\,g^3\,\frac{3(N^2-1)(N^2-4)(N^2+3)}{32\sqrt{2}N^2}
-g^5\,\frac{(N^2-1)(N^2-4)(N^4+2N^2-8)}{128\sqrt{2}N^3}\\
&~~~\,-g^7\,\frac{(N^2-1)(N^2-4)(3 N^6+5 N^4-35 N^2+75)}{12288
   \sqrt{2} N^4}+\cdots~,
\end{aligned}
\label{X3exp}
\end{equation}
while at level $n=4$ we get
\begin{equation}
\begin{aligned}
\cX_{(4)}&=-\,g^4\,\frac{(N^2-1)(N^6+2N^4-18N^2+81)}{96N^3}\\
&~~~\,-g^6
\frac{(N^2-1)(2 N^8+5 N^6-41 N^4+270 N^2-486)}{3072 N^4}\\
&~~~\,-g^8\,\frac{(N^2-1)(2 N^{10}+9 N^8-53 N^6+270 N^4-960 N^2+1710)}{122880 N^5}+\cdots~,
\end{aligned}
\label{X4exp}
\end{equation}
and
\begin{align}
\cX_{(2,2)}&=-\,g^4\,\frac{(N^2-1)(2N^2-3)(N^2+3)}{32 N^2}
-g^6\frac{(N^2-1)(7N^2+27)(N^4-3 N^2+3)}{1536 N^3}\notag
\\
&~~~\,-g^8\,\frac{(N^2-1)(4 N^2+19)(2 N^6-8 N^4+15 N^2-15)}{61440 N^4}+\cdots~.
\label{X22exp}
\end{align}
Multiplying the leading terms in these expansions by $\frac{3\,\zeta(3)\,g^4}{(8\pi^2)^2}$,
we obtain the 2-loop corrections to the amplitudes
$\cA_{\vec{n}}$, whose explicit expressions are collected in Tab.~\ref{tab2} for all operators
up to dimension $n=4$.

\begin{table}[ht]
\begin{center}
{
\begin{tabular}{|c|c|}
\hline
$\vec{n}\phantom{\bigg|}$&$\delta \cA_{\vec{n}}\big|_{\mathrm{2-loop}}$\\
      \hline\hline
$(2)\phantom{\bigg|}$ & $-g^6\frac{\zeta(3)}{(8\pi^2)^2\phantom{\big|}}
\frac{9(N^2-1)(N^2+1)}{8N\phantom{\big|}}$ \\
\hline
$(3)\phantom{\bigg|}$ & $-g^7\frac{\zeta(3)}{(8\pi^2)^2\phantom{\big|}}\frac{9(N^2-1)(N^2-4)(N^2+3)}{32\sqrt{2}N^2\phantom{\big|}}$ \\
\hline
$(4)\phantom{\bigg|}$ & $-g^8\frac{\zeta(3)}{(8\pi^2)^2\phantom{\big|}}
\frac{(N^2-1)(N^6+2N^4-18N^2+81)}{32N^3\phantom{\big|}} $ \\
\hline
$(2,2)\phantom{\bigg|}$ & $-g^8\frac{\zeta(3)}{(8\pi^2)^2\phantom{\big|}}
\frac{3(N^2-1)(2N^2-3)(N^2+3)}{32 N^2\phantom{\big|}}$ \\
\hline
\end{tabular}
}
\end{center}
\caption{The 2-loop contribution to the difference $\delta\cA_{\vec{n}}$ between the $\cN=2$
and the $\cN=4$ amplitudes for operators up to order $n=4$.}
\label{tab2}
\end{table}

It should be clear by now that this procedure can be used to find $\cX_{\vec{n}}$ for any $\vec{n}$, 
and also that it can be straightforwardly generalized to obtain the exact 
expressions of the  corrections with higher transcendentality, like for example $\cY_{\vec{n}}$ 
in (\ref{deltaAnexp}).
Of course, the resulting formulas become longer and longer when one goes higher and higher 
in $n$ or in transcendentality; however this approach, which is essentially based on the use of the 
recursion relations (\ref{rrecursion}), provides a systematic way to obtain exact expressions to any desired
order.

\subsection{The large-$N$ limit}
\label{secn:largeN}

We now study the behavior of the matrix model amplitudes in the planar limit $N\to \infty$
with the 't~Hooft coupling
\begin{equation}
\lambda=g^2 N
\label{lambda}
\end{equation}
kept fixed.
We begin with the $\cN=4$ theory and later turn to the superconformal $\cN=2$ model.

\subsubsection*{The $\cN=4$ theory}
Taking the planar limit of the expectation value of the Wilson loop, from (\ref{hatA0exp}) we get
\begin{equation}
\widehat{\cA}_{(0)}\Big|_{\mathrm{planar}} = 1+\frac{\lambda}{8}+\frac{\lambda^2}{192}
+\frac{\lambda^3}{9216}+\cdots =\frac{2}{\sqrt{\lambda}}\,I_1\big(\sqrt{\lambda}\big)
\label{hatA0planar}
\end{equation}
where $I_n$ is the modified Bessel function of the first kind.
This is a well-known and established result \cite{Erickson:2000af}.

Next, let us consider the amplitude between the Wilson loop and the operator at level $n=2$ given in
(\ref{A21g}). In the planar limit it becomes
\begin{equation}
\widehat{\cA}_{(2)}\Big|_{\mathrm{planar}} = \frac{\lambda}{8}+\frac{\lambda^2}{96}
+\frac{\lambda^3}{3072}+\cdots =I_2\big(\sqrt{\lambda}\big)~.
\label{hatA2planar}
\end{equation}
Also this is a known result \cite{Semenoff:2001xp}.

Proceeding systematically in this way and using the explicit results in the Gaussian matrix model, 
it is not difficult to find the weak-coupling
expansion of the amplitude $\widehat{\cA}_{\vec{n}}$ in the planar limit for a generic operator, and also
to obtain its exact resummation in terms of Bessel functions. Indeed, for 
a generic vector $\vec{n}$ one can show that
\begin{equation}
g^{n-2\ell}\,\widehat{\cA}_{\vec{n}}\Big|_{\mathrm{planar}} 
= \frac{\big(\sqrt{\lambda}\big)^{n-\ell-1}}{2^{\frac{n}{2}+\ell-1}\phantom{\big|}}
\,I_{n-\ell+1}\big(\sqrt{\lambda}\big)\,\prod_{i=1}^\ell n_i
\label{hatAnplanar}
\end{equation}
where $n$ is, as usual, the sum of the components of $\vec{n}$ (see (\ref{defn})), 
while $\ell$ is the number of these components, namely the number of traces that appear in the
corresponding operator. We have verified the validity of this formula by explicitly computing 
the planar limit of the amplitudes between the Wilson loop and all operators up to dimension $n=7$. 
In Tab.~\ref{tab3} we collect our results up to level $n=4$. We point out that
for $\ell=1$, {\it{i.e.}} for the single trace operators, our formula (\ref{hatAnplanar}) agrees 
with the findings of \cite{Semenoff:2001xp}.

\begin{table}[ht]
\begin{center}
{
\begin{tabular}{|c|c|c|}
\hline
\multirow{2}{*}{$\vec{n}$} & Expansion of
& ~~Exact expression of ~~\\
      & ~~$g^{n-2\ell}\,\widehat{\cA}_{\vec{n}}\big|_{\mathrm{planar}}
\phantom{\Big|}$ &~~$g^{n-2\ell}\,\widehat{\cA}_{\vec{n}}\big|_{\mathrm{planar}}
\phantom{\Big|}$\\
      \hline\hline
$(2)$ & $
\frac{\lambda}{8}
+ \frac{\lambda^2}{96}
+ \frac{\lambda^3}{3072} 
+\cdots\phantom{\bigg|}$ 
& $I_2\big(\sqrt{\lambda}\big)$
\\ \hline
$(3)$ & $
\!\frac{\lambda^2}{32\sqrt{2}}+\frac{\lambda^3}{512\sqrt{2}}
+\frac{\lambda^4}{20480\sqrt{2}}
+\cdots\phantom{\bigg|}$ 
& $\frac{3\sqrt{\lambda}}{2\sqrt{2}}\,I_3\big(\sqrt{\lambda}\big)$
\\ \hline
$(4)$ & $
\frac{\lambda^3}{384}+\frac{\lambda^4}{7680}
+\frac{\lambda^5}{368640}
+\cdots\phantom{\bigg|}$ 
& $\lambda\,I_4\big(\sqrt{\lambda}\big)$
\\ \hline
$(2,2)$ & $
\frac{\lambda^2}{96}+ \frac{\lambda^3}{1536} +\frac{\lambda^4}{61440}
+\cdots\phantom{\bigg|}$ 
&$\frac{\sqrt{\lambda}}{2}\,I_3\big(\sqrt{\lambda}\big)$
\\ \hline
\end{tabular}
}
\end{center}
\caption{Results for the $\mathcal{N}=4$ matrix model in the planar limit. As explained in the text, $n$ is the 
sum of the components of $\vec{n}$ while $\ell$ is the number of these components.}
\label{tab3}
\end{table}

\subsubsection*{The $\cN=2$ superconformal theory}
Multiplying (\ref{X0}) by $\frac{3\,\zeta(3)\,g^4}{(8\pi^2)^2}$ and then taking the large $N$ limit, it is
straightforward to obtain\,\footnote{See also \cite{Sysoeva:2017fhr}.}
\begin{equation}
\delta{\cA}_{(0)}\Big|_{\mathrm{planar}}=-\frac{3\,\zeta(3)\,\lambda^2}{(8\pi^2)^2}\Big(
\frac{\lambda}{8}+\frac{2\lambda^2}{192}
+\frac{3\lambda^3}{9216}+\cdots\Big)+\cdots=-\frac{3\,\zeta(3)\,\lambda^2}{(8\pi^2)^2}\,I_2\big(
\sqrt{\lambda}\big)+\cdots
\label{deltaA0planar}
\end{equation}
where the last ellipses stand for terms of higher transcendentality.

In a similar way, from (\ref{X2b}) we easily get
\begin{equation}
\delta{\cA}_{(2)}\Big|_{\mathrm{planar}}=-\frac{3\,\zeta(3)\,\lambda^2}{(8\pi^2)^2}\Big(
\frac{3\lambda}{8}+\frac{4\lambda^2}{96}
+\frac{5\lambda^3}{3072}+\cdots\Big)+\cdots~.
\label{deltaA2planar}
\end{equation}
It is interesting to observe that if one compares this expression with the expansion of
the planar limit of the $\cN=4$ amplitude $\widehat{\cA}_{(2)}$ given in
(\ref{hatA2planar}), one sees that each term of the latter proportional to $\lambda^k$
gets multiplied by
\begin{equation}
-\frac{3\,\zeta(3)\,\lambda^2}{(8\pi^2)^2}\,(k+2)~.
\label{shift2}
\end{equation}
As we will see in Section~\ref{sec:pert}, this fact has a simple and nice diagrammatic interpretation. The
expansion (\ref{deltaA2planar}) can be resummed in terms of modified Bessel functions as
follows
\begin{equation}
\delta{\cA}_{(2)}\Big|_{\mathrm{planar}}=
-\frac{3\,\zeta(3)\,(\sqrt{\lambda})^5}{2(8\pi^2)^2\phantom{\big|}}\Big(I_1\big(\sqrt{\lambda}\big)
+\frac{2}{\sqrt{\lambda}} I_2\big(\sqrt{\lambda}\big)\Big)+\cdots~.
\label{deltaA2planar1}
\end{equation}
Taking into account the different normalization of the operator $\cO_{(2)}(a)$ we have used, 
our result agrees with \cite{Rodriguez-Gomez:2016cem}.

Proceeding in this way and using (\ref{X3exp})--(\ref{X22exp}), 
it is not difficult to obtain the weak-coupling expansions of $\delta\cA_{(3)}$, $\delta\cA_{(4)}$ and 
$\delta\cA_{(2,2)}$ in the planar limit, and eventually their exact expressions.
In Tab.~\ref{tab4} we have collected our findings for the terms proportional 
to $\zeta(3)$ in $\delta\cA_{\vec{n}}$ for all operators up to dimension $n=4$.

\begin{table}[ht]
\begin{center}
{
\begin{tabular}{|c|c|c|}
\hline
\multirow{2}{*}{$\vec{n}$} & Expansion of the $\zeta(3)$-term of
& ~~Exact expression of the $\zeta(3)$-term of ~~\\
      & ~~$g^{n-2\ell}\,\delta{\cA}_{\vec{n}}\big|_{\mathrm{planar}}
\phantom{\Big|}$ &~~$g^{n-2\ell}\,\delta{\cA}_{\vec{n}}\big|_{\mathrm{planar}}
\phantom{\Big|}$\\
      \hline\hline
$(2)$ & $-\frac{3\,\zeta(3)\,\lambda^2}{(8\pi^2)^2\phantom{\big|}}\Big(
\frac{3\lambda}{8}
+ \frac{4\lambda^2}{96}
+ \frac{5\lambda^3}{3072} +\cdots\!\Big)\!
\!\!\phantom{\bigg|}$ 
& $-\frac{3\,\zeta(3)\,(\sqrt{\lambda})^5}{2(8\pi^2)^2\phantom{\big|}}\Big(I_1\big(\sqrt{\lambda}\big)
+\frac{2}{\sqrt{\lambda}} I_2\big(\sqrt{\lambda}\big)\Big)$
\\ \hline
$(3)$ & $\!\!
-\frac{3\,\zeta(3)\,\lambda^2}{(8\pi^2)^2\phantom{\big|}}\Big(\frac{3\lambda^2}{32\sqrt{2}}\!
+\!\frac{4\lambda^3}{512\sqrt{2}}
+\frac{5\lambda^4}{20480\sqrt{2}}\!+\cdots\!\Big)\!\!
\!\!\phantom{\bigg|}$ 
& $-\frac{9\,\zeta(3)\,\lambda^3}{4\sqrt{2}(8\pi^2)^2\phantom{\big|}}\,I_2\big(\sqrt{\lambda}\big)$
\\ \hline
$(4)$ & $
-\frac{3\,\zeta(3)\,\lambda^2}{(8\pi^2)^2\phantom{\big|}}\Big(\frac{4\lambda^3}{384}+\frac{5\lambda^4}{7680}
+\frac{6\lambda^5}{368640}
+\cdots\!\Big)\phantom{\bigg|}$ 
& $-\frac{3\,\zeta(3)\,(\sqrt{\lambda})^7}{2(8\pi^2)^2\phantom{\big|}}\,I_3\big(\sqrt{\lambda}\big)$
\\ \hline
$(2,2)$ & $
-\frac{3\,\zeta(3)\,\lambda^2}{(8\pi^2)^2\phantom{\big|}}\Big(\frac{6\lambda^2}{96}+ \frac{7\lambda^3}{1536} +\frac{8\lambda^4}{61440}
+\cdots\!\Big)\phantom{\bigg|}$ 
&$-\frac{3\,\zeta(3)\,\lambda^3}{4(8\pi^2)^2\phantom{\big|}}\Big(I_2\big(\sqrt{\lambda}\big)
+\frac{6}{\sqrt{\lambda}} I_3\big(\sqrt{\lambda}\big)\Big)$
\\ \hline
\end{tabular}
}
\end{center}
\caption{Results for the $\mathcal{N}=2$ superconformal matrix model in the planar limit. As before, $n$ is 
the sum of the components of $\vec{n}$ while $\ell$ is their number.}
\label{tab4}
\end{table}

{From} these explicit results it is possible to infer the following general formula
\begin{align}
g^{n-2\ell}\,\delta{\cA}_{\vec{n}}\Big|_{\mathrm{planar}}
&=-\frac{3\,\zeta(3)}{(8\pi^2)^2}\,
\frac{\big(\sqrt{\lambda}\big)^{n-\ell+4}}{2^{\frac{n}{2}+\ell}\phantom{\big|}}\left\{\bigg[
I_{n-\ell}\big(\sqrt{\lambda}\big)
+\frac{2(\ell-1)}{\sqrt{\lambda}}\,I_{n-\ell+1}\big(\sqrt{\lambda}\big)\bigg]\prod_{k=1}^\ell n_k
\right.
\notag\\
&\left.~~~+\bigg(\sum_{i=1}^\ell \delta_{n_i,2}\bigg)\frac{2}{\sqrt{\lambda}}\,
I_{n-\ell+1}\big(\sqrt{\lambda}\big)\prod_{k=1}^\ell n_k\right\}+\cdots
\label{deltaAnexact}
\end{align}
which we have verified in all cases up to $n=7$. We observe that there is a contribution, represented by 
the second line above, which occurs only when the operator $\cO_{\vec{n}}(a)$ contains at least
a factor of the type $\tr a^2$. This fact has a precise diagrammatic counterpart, as we will see
in the next section.

Comparing the two exact expressions (\ref{deltaAnexact}) and (\ref{hatAnplanar}) and using the 
properties of the modified Bessel functions, it is not difficult to realize that
\begin{equation}
g^{n-2\ell}\,\delta\cA_{\vec{n}}\Big|_{\mathrm{planar}}= -\frac{3\,\zeta(3)\,\lambda^2}{(8\pi^2)^2}\,
\bigg(\lambda\,\frac{\partial}{\partial\lambda}+\ell+
\sum_{i=1}^\ell\delta_{n_i,2}\bigg)\bigg(g^{n-2\ell}\,\widehat{\cA}_{\vec{n}}\Big|_{\mathrm{planar}}\bigg)+\cdots
\label{deltaAnhatAn}
\end{equation}
where, as usual, the ellipses stand for terms of higher transcendentality. Such a relation implies
that if we multiply each term $\lambda^k$ in the weak-coupling expansion
of $g^{n-2\ell}\,\widehat{\cA}_{\vec{n}}\big|_{\mathrm{planar}}$ by
\begin{equation}
-\frac{3\,\zeta(3)\,\lambda^2}{(8\pi^2)^2}\,\Big(k+\ell+\sum_{i=1}^\ell \delta_{n_i,2}\Big)~,
\label{shiftn}
\end{equation}
then we obtain the expansion of the $\zeta(3)$-correction of the corresponding $\cN=2$ planar amplitude
$g^{n-2\ell}\,\delta\cA_{\vec{n}}\big|_{\mathrm{planar}}$. Also this formula, which generalizes 
(\ref{shift2}) to any $\vec{n}$, has a simple and nice interpretation in terms of field theory diagrams, 
as we will see in the next section. 

\section{Perturbative checks in field theory}
\label{sec:pert}
We now consider the direct field theory computation of the expectation values of chiral operators 
with a circular BPS Wilson loop in a superconformal $\mathcal{N}=2$ theory defined 
on $\mathbb{R}^4$.

As explained in Section~\ref{sec:WL}, conformal invariance implies that all information about these 
expectation values is contained in the amplitudes $A_{\vec{n}}$ defined in (\ref{Ang}). 
The conjecture we want to test is that these amplitudes match the corresponding ones $\cA_{\vec{n}}$
in the matrix model that we introduced in (\ref{Angdef}), namely we want to show that
\begin{equation}
\label{AisCA}
A_{\vec n} \,=\, \cA_{\vec n}~.
\end{equation}
The diagrammatic evaluation in field theory of the correlators $A_{\vec n}$ beyond tree-level is 
in general quite complicated. However, it becomes tractable if one only computes 
the difference between the $\mathcal{N}=2$ result and the one we would have in the $\mathcal{N}=4$ theory. This is the same strategy utilized in \cite{Andree:2010na} to check the matrix model expression (\ref{deltaA02loop}) for the $\mathcal{N}=2$ Wilson loop itself, as well as in 
\cite{Billo:2017glv} to compute chiral-antichiral two-point functions in absence of Wilson loops.
We now briefly recall the main steps of this approach.
 
We first split the $\mathcal{N}=2$ action as:
\begin{equation}
\label{S2action}
S^{(N_f)}_{\mathcal{N}=2}= S_{\mathrm{gauge}}+S_Q~,
\end{equation}
separating the pure gauge term, $S_{\mathrm{gauge}}$, with the $\mathcal{N}=2$ vector 
multiplet from the matter term, $S_Q$, which contains $N_f$ hypermultiplets $Q$ in the fundamental representation of the gauge group. Then, we view the $\mathcal{N}=4$ vector multiplet as 
a combination of a $\mathcal{N}=2$ vector with an adjoint $\mathcal{N}=2$ hypermultiplet $H$; 
in this way the $\mathcal{N}=4$ SYM action can be written as: 
\begin{equation}
\label{S4action}
S_{\mathcal{N}=4}= S_{\mathrm{gauge}}+S_H~,
\end{equation}
so that
\begin{equation}
S^{(N_f)}_{\mathcal{N}=2}= S_{\mathcal{N}=4}+S_Q-S_H~.
\label{S2action1}
\end{equation}
All terms in the right hand side of (\ref{S2action1}) have a well-established $\mathcal{N}=1$ 
superfield formulation, which allows us to easily write down the Feynman rules in configuration space. 
For this we refer to Section 3.1 of \cite{Billo:2017glv}, whose notations and conventions we
consistently use in the following.

{From} (\ref{S2action1}) we deduce that any correlator $A_{\vec{n}}$ of  the $\mathcal{N}=2$ theory 
can be written as:
\begin{equation}
A_{\vec n}= \widehat A_{\vec n} + A_{\vec n,Q} - A_{\vec n,H} 
\end{equation}
where $\widehat A_{\vec n}$ is the correlator in the $\cN =4$ theory, while 
$A_{\vec n,H}$ and $A_{\vec n,Q}$ are the contributions from diagrams in which the adjoint hypermultiplet $H$ and the fundamental hypermultiplets $Q$ run in the internal lines. 
Therefore the difference between the $\cN=2$ and the $\cN=4$ amplitudes is
\begin{equation}
\delta A_{\vec{n}}= A_{\vec{n}}-\widehat{A}_{\vec{n}}= A_{\vec n,Q} - A_{\vec n,H} ~.
\end{equation}
Performing this diagrammatic difference in the perturbative field theory computations leads to remarkable simplifications, since all diagrams without $Q$ or $H$ internal lines do not need to be considered.

Starting from this set up, what we shall check, up to two loops, is in fact the following equality:
\begin{equation}
\label{checkN2N4}
\delta A_{\vec n} = \delta \cA_{\vec n}~, 
\end{equation}
where $\delta \cA_{\vec n}$ is the difference between the 
$\cN=2$ and $\cN=4$ matrix model results introduced in (\ref{diffAn}). 

\subsection{Tree-level}
\label{subsec:tree}
At the lowest order in $g$ the $\cN=2$ and $\cN=4$ amplitudes coincide:
\begin{equation}
\label{Atree}
A_{\vec n}\Big|_{\mathrm{tree-level}} = 
\widehat A_{\vec n}\Big|_{\mathrm{tree-level}}~;
\end{equation}
in other words, 
\begin{equation}
\delta A_{\vec n}\Big|_{\mathrm{tree-level}} = 0~.
\end{equation}
Also in the matrix model this difference vanishes at the lowest order, see (\ref{deltaAntree1loop}).
Thus, the equality (\ref{checkN2N4}) is satisfied at tree level. 

\begin{figure}[ht]
\begin{center}
\begingroup%
  \makeatletter%
  \providecommand\color[2][]{%
    \errmessage{(Inkscape) Color is used for the text in Inkscape, but the package 'color.sty' is not loaded}%
    \renewcommand\color[2][]{}%
  }%
  \providecommand\transparent[1]{%
    \errmessage{(Inkscape) Transparency is used (non-zero) for the text in Inkscape, but the package 'transparent.sty' is not loaded}%
    \renewcommand\transparent[1]{}%
  }%
  \providecommand\rotatebox[2]{#2}%
  \ifx\svgwidth\undefined%
    \setlength{\unitlength}{200bp}%
    \ifx\svgscale\undefined%
      \relax%
    \else%
      \setlength{\unitlength}{\unitlength * \real{\svgscale}}%
    \fi%
  \else%
    \setlength{\unitlength}{\svgwidth}%
  \fi%
  \global\let\svgwidth\undefined%
  \global\let\svgscale\undefined%
  \makeatother%
  \begin{picture}(1,0.63419123)%
    \put(0,0){\includegraphics[width=\unitlength,page=1]{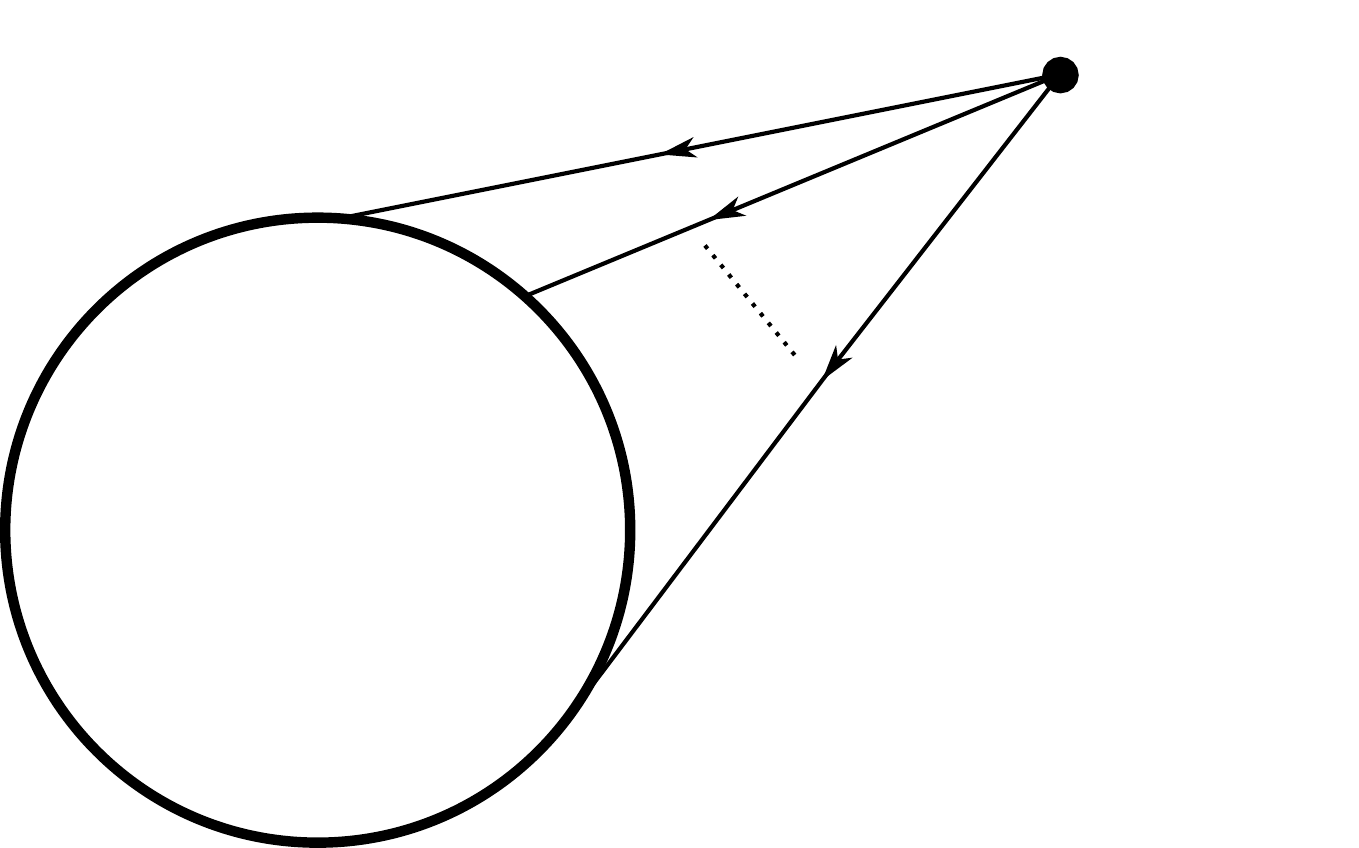}}%
    \put(0.71542891,0.61051217){\color[rgb]{0,0,0}\makebox(0,0)[lb]{\smash{$O_{\vec{n}}(x)$}}}%
    \put(0.05320224,0.12273485){\color[rgb]{0,0,0}\makebox(0,0)[lb]{\smash{$W(C)$}}}%
  \end{picture}%
\endgroup%
\end{center}
\caption{At the lowest perturbative order, the operator $O_{\vec n}(x)$ is connected to the Wilson loop by $n$ scalar propagators. Exploiting conformal invariance, we can place the operator in the origin, {\it{i.e.}}
in the center of the Wilson loop. Nevertheless, in this and in the following pictures we will continue to 
place it outside the loop to avoid graphical clutter.}
\label{fig:WOntree}
\end{figure}

Actually, in this case it is easy (and also convenient for later purposes) 
to check directly the validity of (\ref{AisCA}). 
Performing this check is helpful also to establish some facts that will be useful at higher orders; 
in particular, the way the path-ordered integration over the Wilson loop simplifies in the tree-level case 
will be exploited also in the two-loop order computations. Thus, for later convenience
we briefly show some details.
At the lowest order in $g$, the $n$ chiral fields $\varphi$ appearing
in the operator $O_{\vec n}$ must be contracted with the $n$ antichiral fields
present in the term of order $n$ in the expansion on the Wilson loop operator (\ref{WLdef2}).
This is represented by the diagram in Fig.~\ref{fig:WOntree}. 

Thus, we have
\begin{equation}
\label{cosW2}
\big\langle\, W(C)\,O_{\vec n}(0) \,\big\rangle\Big|_{\mathrm{tree-level}} =\frac{1}{N}\,\frac{g^{n}}{n!}\,
\Big\langle\, \mathcal{P}\,\tr\Big({\prod_{i=1}^n} \oint_{C}d\tau_{i}\,
\frac{R}{\sqrt{2}}\,\bar{\varphi}(x_i)\Big) O_{\vec n}(0) \,\Big\rangle 
\end{equation}
where we have denoted by $x_i = x(\tau_i)$ the positions along the Wilson loop $C$.
Using (\ref{OnR}), we rewrite this expression as
\begin{align}
\label{cosW3}
\big\langle\,W(C)\,O_{\vec n}(0)\,\big\rangle\Big|_{\mathrm{tree-level}} &= 
\frac{1}{N}\,\frac{g^n \,R^n}{2^{\frac n2}\,n!}\,
\cP\, {\prod_{i=1}^n} \oint_{C}d\tau_{i}\,
\tr \big(T^{a_1} \cdots  T^{a_n} \big) \,R^{b_1\ldots b_n}_{\vec{n}}\nonumber\\
& ~~~\times \big\langle\,\bar{\varphi}^{a_1}(x_1)\cdots \bar{\varphi}^{a_n}(x_n)
\,\varphi^{b_1}(0) \cdots \varphi^{b_n}(0)\,\big\rangle~.
\end{align}
The vacuum expectation value in the second line above is computed using the free scalar propagator
\begin{equation}
\label{scalprop}
\big\langle\,\bar\varphi^{a}(x_i)\,\varphi^{b}(0)\,\big\rangle 
= \frac{\delta^{ab}}{4\pi^2\,x_i^2}= \frac{\delta^{ab}}{4\pi^2 R^2}
\end{equation} 
where we have exploited the fact that $x_i = x(\tau_i)$ belongs to the circle $C$ of radius $R$ and thus can
be parameterized as in (\ref{circle}). In view of this, 
when we apply Wick's theorem in (\ref{cosW3}) we obtain an integrand that  
does not depend on the variables $\tau_i$. 
The path ordering becomes therefore irrelevant and, from the integration over $\tau_i$, we 
simply get a factor of $(2\pi)^n$. Moreover the $n!$ different contraction patterns all yield the same expression, due to the symmetry of the tensor $R_{\vec n}$. 
Taking all this into account, we get
\begin{equation}
\label{cosW4}
\big\langle\,W(C)\,O_{\vec n}(0)\,\big\rangle\Big|_{\mathrm{tree-level}} 
= \frac{1}{N}\,\frac{g^n}{2^{\frac n2}}\,
\frac{1}{(2\pi R)^n}\,R_{\vec{n}}^{\,b_1\dots b_n}\,\tr \big(T_{b^1}\dots T^{b_n}\big)~,
\end{equation}
which implies that
\begin{equation}
\label{Antree0}
A_{\vec n}\Big|_{\mathrm{tree-level}} \,=\, 
\widehat A_{\vec n}\Big|_{\mathrm{tree-level}} 
=\frac{g^n}{N\,2^{\frac{n}{2}}}\,R_{\vec{n}}^{\,b_1\dots b_n}\,\tr \big(T^{b_1}\dots T^{b_n}\big)~,
\end{equation}
in full agreement with the matrix model result (\ref{Antree}).

\subsection{Loop corrections}
\label{subsec:2loop} 
\begin{figure}[ht]
\begin{center}
\vspace{0.3cm}
\begingroup%
  \makeatletter%
  \providecommand\color[2][]{%
    \errmessage{(Inkscape) Color is used for the text in Inkscape, but the package 'color.sty' is not loaded}%
    \renewcommand\color[2][]{}%
  }%
  \providecommand\transparent[1]{%
    \errmessage{(Inkscape) Transparency is used (non-zero) for the text in Inkscape, but the package 'transparent.sty' is not loaded}%
    \renewcommand\transparent[1]{}%
  }%
  \providecommand\rotatebox[2]{#2}%
  \ifx\svgwidth\undefined%
    \setlength{\unitlength}{320bp}%
    \ifx\svgscale\undefined%
      \relax%
    \else%
      \setlength{\unitlength}{\unitlength * \real{\svgscale}}%
    \fi%
  \else%
    \setlength{\unitlength}{\svgwidth}%
  \fi%
  \global\let\svgwidth\undefined%
  \global\let\svgscale\undefined%
  \makeatother%
  \begin{picture}(1,0.65427991)%
    \put(0,0){\includegraphics[width=\unitlength,page=1]{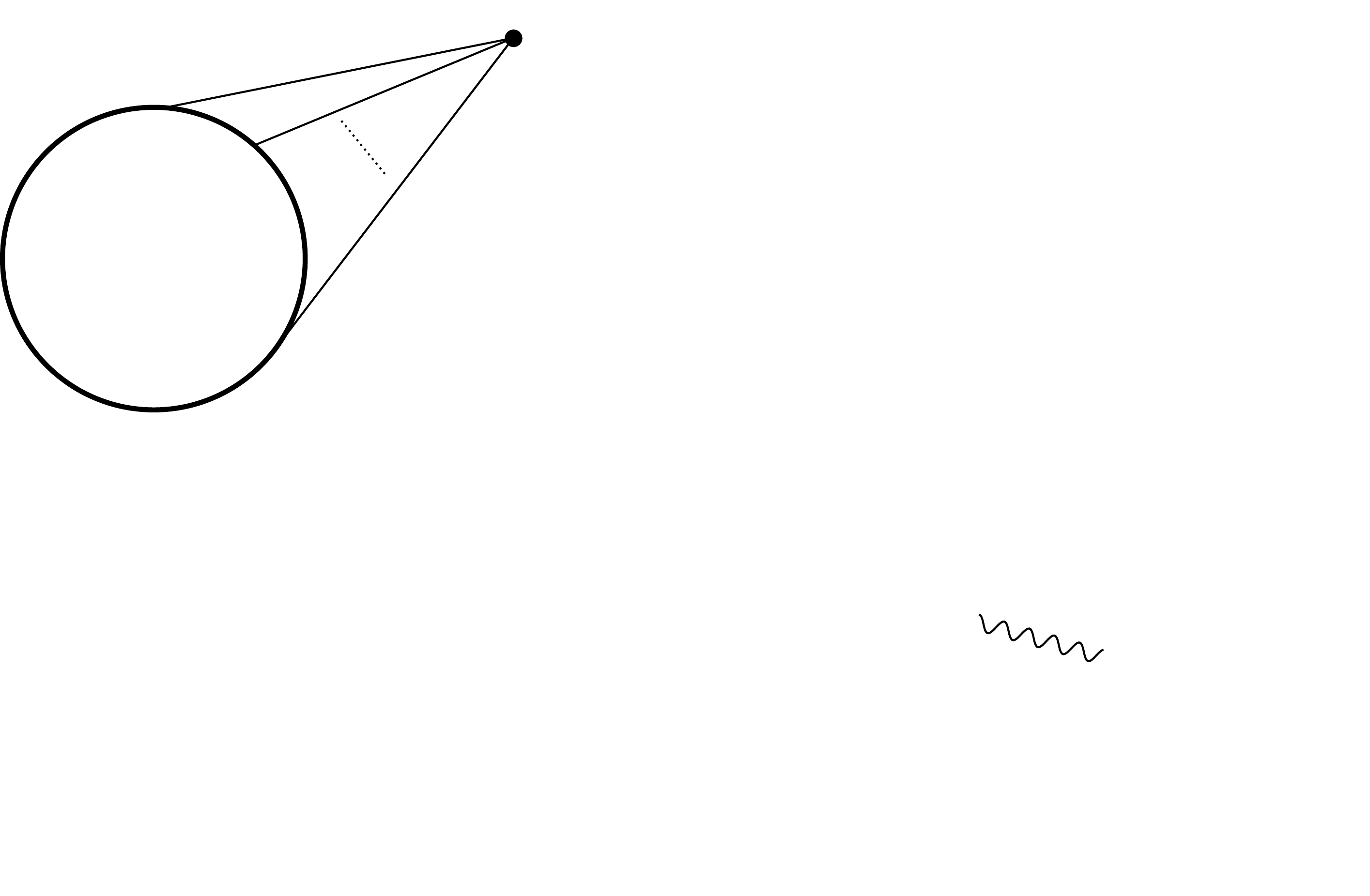}}%
    \put(0.32195035,0.64164467){\color[rgb]{0,0,0}\makebox(0,0)[lb]{\smash{$O_{\vec n}(x)$}}}%
    \put(0.04447309,0.38718835){\color[rgb]{0,0,0}\makebox(0,0)[lb]{\smash{\textbf{$W(C)$}}}}%
    \put(0,0){\includegraphics[width=\unitlength,page=2]{WOnN4.pdf}}%
    \put(0.55837637,0.38718835){\color[rgb]{0,0,0}\makebox(0,0)[lb]{\smash{\textbf{$W(C)$}}}}%
    \put(0,0){\includegraphics[width=\unitlength,page=3]{WOnN4.pdf}}%
    \put(0,0){\includegraphics[width=\unitlength,page=4]{WOnN4.pdf}}%
    \put(0.04447309,0.03415112){\color[rgb]{0,0,0}\makebox(0,0)[lb]{\smash{\textbf{$W(C)$}}}}%
    \put(0,0){\includegraphics[width=\unitlength,page=5]{WOnN4.pdf}}%
    \put(0.55837637,0.03415112){\color[rgb]{0,0,0}\makebox(0,0)[lb]{\smash{\textbf{$W(C)$}}}}%
    \put(0.83959469,0.64308611){\color[rgb]{0,0,0}\makebox(0,0)[lb]{\smash{$O_{\vec n}(x)$}}}%
    \put(0.32819705,0.29132388){\color[rgb]{0,0,0}\makebox(0,0)[lb]{\smash{$O_{\vec n}(x)$}}}%
    \put(0.82715765,0.29158373){\color[rgb]{0,0,0}\makebox(0,0)[lb]{\smash{$O_{\vec n}(x)$}}}%
  \end{picture}%
\endgroup%
\end{center}
\caption{Diagrams which do not contain interaction vertices including $H$ or $Q$ hypermultiplets and
which therefore vanish in the difference between the $\cN=2$ and the $\cN=4$ theory. Here there are some examples of diagrams which appear at order $g^2$ with respect to the tree-level amplitude
$A_{\vec{n}}$, but vanish in the difference $\delta A_{\vec{n}}$.}
\label{fig:WOnN4}
\end{figure}

At higher orders in $g$ we concentrate on the difference $\delta A_{\vec{n}}$. As we already
pointed out, the number of diagrams which contribute to this difference
is massively reduced. For example, all diagrams represented in Fig.~\ref{fig:WOnN4} yield a $g^2$
correction with respect to the tree-level amplitude $A_{\vec{n}}$ but they should not be considered
in the computation of $\delta A_{\vec{n}}$ since they do not contain internal lines with $H$ or $Q$ hypermultiplets.

\subsubsection*{One loop}
\begin{figure}[ht]
\begin{center}
\begingroup%
  \makeatletter%
  \providecommand\color[2][]{%
    \errmessage{(Inkscape) Color is used for the text in Inkscape, but the package 'color.sty' is not loaded}%
    \renewcommand\color[2][]{}%
  }%
  \providecommand\transparent[1]{%
    \errmessage{(Inkscape) Transparency is used (non-zero) for the text in Inkscape, but the package 'transparent.sty' is not loaded}%
    \renewcommand\transparent[1]{}%
  }%
  \providecommand\rotatebox[2]{#2}%
  \ifx\svgwidth\undefined%
    \setlength{\unitlength}{200bp}%
    \ifx\svgscale\undefined%
      \relax%
    \else%
      \setlength{\unitlength}{\unitlength * \real{\svgscale}}%
    \fi%
  \else%
    \setlength{\unitlength}{\svgwidth}%
  \fi%
  \global\let\svgwidth\undefined%
  \global\let\svgscale\undefined%
  \makeatother%
  \begin{picture}(1,0.63419123)%
    \put(0,0){\includegraphics[width=\unitlength,page=1]{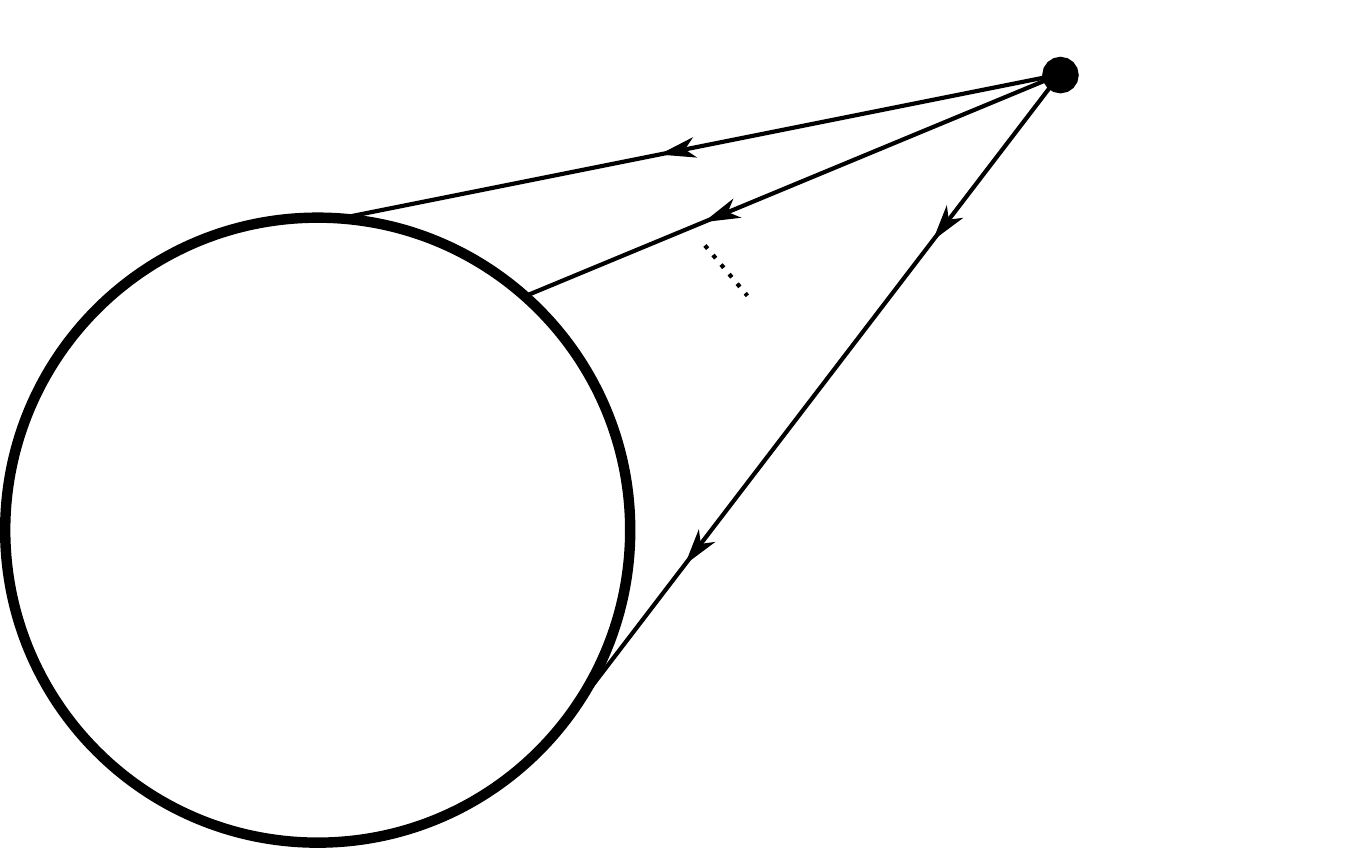}}%
    \put(0.70542891,0.60051217){\color[rgb]{0,0,0}\makebox(0,0)[lb]{\smash{$O_{\vec{n}}(x)$}}}%
    \put(0.05320224,0.12273485){\color[rgb]{0,0,0}\makebox(0,0)[lb]{\smash{$W(C)$}}}%
    \put(0,0){\includegraphics[width=\unitlength,page=2]{WOn1loop.pdf}}%
    \put(0.59002566,0.32527747){\color[rgb]{0,0,0}\makebox(0,0)[lb]{\smash{$1$}}}%
  \end{picture}%
\endgroup%
\end{center}
\caption{The only diagrams that yield a $g^2$ correction to the tree-level amplitude $A_{\vec{n}}$
and contain $Q$ and $H$ hypermultiplets arise from the one-loop correction of the external 
scalar propagators.}
\label{fig:WOn1loop}
\end{figure}

\begin{figure}[ht]
\begin{center}
\vspace{0.3cm}
\begingroup%
  \makeatletter%
  \providecommand\color[2][]{%
    \errmessage{(Inkscape) Color is used for the text in Inkscape, but the package 'color.sty' is not loaded}%
    \renewcommand\color[2][]{}%
  }%
  \providecommand\transparent[1]{%
    \errmessage{(Inkscape) Transparency is used (non-zero) for the text in Inkscape, but the package 'transparent.sty' is not loaded}%
    \renewcommand\transparent[1]{}%
  }%
  \providecommand\rotatebox[2]{#2}%
  \ifx\svgwidth\undefined%
    \setlength{\unitlength}{300bp}%
    \ifx\svgscale\undefined%
      \relax%
    \else%
      \setlength{\unitlength}{\unitlength * \real{\svgscale}}%
    \fi%
  \else%
    \setlength{\unitlength}{\svgwidth}%
  \fi%
  \global\let\svgwidth\undefined%
  \global\let\svgscale\undefined%
  \makeatother%
  \begin{picture}(1,0.29014587)%
    \put(0,0){\includegraphics[width=\unitlength,page=1]{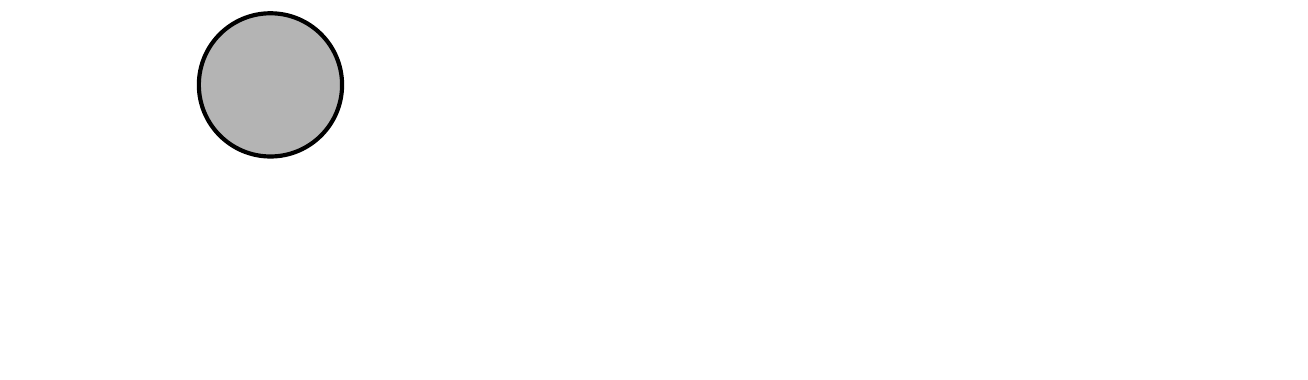}}%
    \put(0.1950383,0.21774438){\color[rgb]{0,0,0}\makebox(0,0)[lb]{\smash{$1$}}}%
    \put(0,0){\includegraphics[width=\unitlength,page=2]{scalprop1l.pdf}}%
    \put(0.47194395,0.21494414){\color[rgb]{0,0,0}\makebox(0,0)[lb]{\smash{\textbf{$=$}}}}%
    \put(0,0){\includegraphics[width=\unitlength,page=3]{scalprop1l.pdf}}%
    \put(0.51767701,0.054211){\color[rgb]{0,0,0}\makebox(0,0)[lb]{\smash{\textbf{$-$}}}}%
    \put(0,0){\includegraphics[width=\unitlength,page=4]{scalprop1l.pdf}}%
 \put(-0.00157339,0.24021085){\color[rgb]{0,0,0}\makebox(0,0)[lb]{\smash{\textbf{$b$}}}}%
    \put(0.38548505,0.24021085){\color[rgb]{0,0,0}\makebox(0,0)[lb]{\smash{\textbf{$a$}}}}%
    \put(0.58504553,0.24021085){\color[rgb]{0,0,0}\makebox(0,0)[lb]{\smash{\textbf{$b$}}}}%
    \put(0.96015877,0.24021085){\color[rgb]{0,0,0}\makebox(0,0)[lb]{\smash{\textbf{$a$}}}}%
    \put(0.58504553,0.07781992){\color[rgb]{0,0,0}\makebox(0,0)[lb]{\smash{\textbf{$b$}}}}%
    \put(0.96015877,0.08156437){\color[rgb]{0,0,0}\makebox(0,0)[lb]{\smash{\textbf{$a$}}}}%
  \end{picture}%
\endgroup%
\end{center}
\caption{The one-loop correction to the scalar propagator. 
The first diagram on the right hand side is the $Q$-contribution due the fundamental hypermultiplets;
the second diagram is the $H$-contribution due the adjoint hypermultiplet and so it comes with a minus sign.}
\label{fig:scalprop1l}
\end{figure}

It is easy to see that in the $\cN=2$ superconformal theory there are no corrections of 
order $g^2$ with respect to the tree-level result. 
In fact, at this order the only possible diagrams containing $H$ and $Q$ hypermultiplets arise 
from the one-loop correction of the external scalar propagators as shown in Fig.~\ref{fig:WOn1loop}. 
This one-loop correction is due to the two diagrams represented in Fig.~\ref{fig:scalprop1l}.
Using the Feynman rules and conventions spelled out in detail in \cite{Billo:2017glv}, one can easily see
that the sum of these two diagrams is proportional to 
\begin{equation}
N_{f}\, \tr \big(T^{b} T^{a}\big) -\big(\ii\,f^{bcd}\big)\,\big(\ii\,f^{adc}\big) = \Big(\frac{N_f}{2} - N\Big)\delta^{ab}~, 
\label{color1loop}
\end{equation}
which vanishes for $N_f=2N$. Therefore, in the superconformal $\cN=2$ theory we have
\begin{equation}
\label{delta1loop}
\delta A_{\vec n}\Big|_{\rm 1-loop} = 0~,
\end{equation}
in full agreement with the matrix model result (see (\ref{deltaAntree1loop})).

\subsubsection*{Two loops}
\begin{figure}[ht]
\vspace{0.3cm}
\begin{center}
\begingroup%
  \makeatletter%
  \providecommand\color[2][]{%
    \errmessage{(Inkscape) Color is used for the text in Inkscape, but the package 'color.sty' is not loaded}%
    \renewcommand\color[2][]{}%
  }%
  \providecommand\transparent[1]{%
    \errmessage{(Inkscape) Transparency is used (non-zero) for the text in Inkscape, but the package 'transparent.sty' is not loaded}%
    \renewcommand\transparent[1]{}%
  }%
  \providecommand\rotatebox[2]{#2}%
  \ifx\svgwidth\undefined%
    \setlength{\unitlength}{450bp}%
    \ifx\svgscale\undefined%
      \relax%
    \else%
      \setlength{\unitlength}{\unitlength * \real{\svgscale}}%
    \fi%
  \else%
    \setlength{\unitlength}{\svgwidth}%
  \fi%
  \global\let\svgwidth\undefined%
  \global\let\svgscale\undefined%
  \makeatother%
  \begin{picture}(1,0.44183123)%
    \put(0,0){\includegraphics[width=\unitlength,page=1]{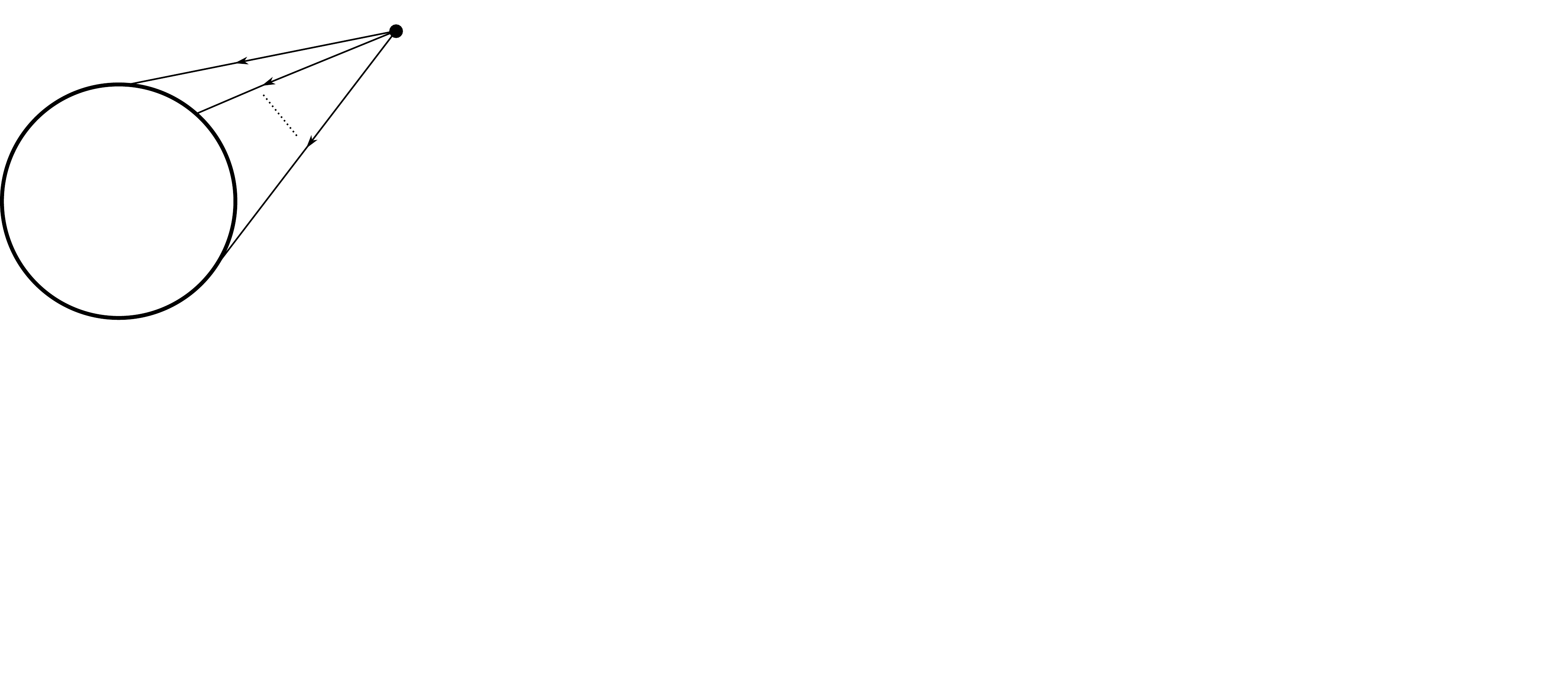}}%
    \put(0.2125034,0.43428981){\color[rgb]{0,0,0}\makebox(0,0)[lb]{\smash{$O_{\vec n}(x)$}}}%
    \put(0.02996213,0.26085401){\color[rgb]{0,0,0}\makebox(0,0)[lb]{\smash{\textbf{$W(C)$}}}}%
    \put(0.17396213,0.26085401){\color[rgb]{0,0,0}\makebox(0,0)[lb]{\smash{\textbf{(a)}}}}%
    \put(0,0){\includegraphics[width=\unitlength,page=2]{WOn1loopg4.pdf}}%
    \put(0,0){\includegraphics[width=\unitlength,page=3]{WOn1loopg4.pdf}}%
    \put(0.08027437,0.34845139){\color[rgb]{0,0,0}\makebox(0,0)[lb]{\smash{$1$}}}%
    \put(0,0){\includegraphics[width=\unitlength,page=4]{WOn1loopg4.pdf}}%
    \put(0.36508569,0.26085401){\color[rgb]{0,0,0}\makebox(0,0)[lb]{\smash{\textbf{$W(C)$}}}}%
    \put(0.51008569,0.26085401){\color[rgb]{0,0,0}\makebox(0,0)[lb]{\smash{\textbf{(b)}}}}%
    \put(0,0){\includegraphics[width=\unitlength,page=5]{WOn1loopg4.pdf}}%
    \put(0,0){\includegraphics[width=\unitlength,page=6]{WOn1loopg4.pdf}}%
    \put(0.41778737,0.31331398){\color[rgb]{0,0,0}\makebox(0,0)[lb]{\smash{$1$}}}%
    \put(0.5514948,0.43428981){\color[rgb]{0,0,0}\makebox(0,0)[lb]{\smash{$O_{\vec n}(x)$}}}%
    \put(0,0){\includegraphics[width=\unitlength,page=7]{WOn1loopg4.pdf}}%
    \put(0.02996213,0.02300807){\color[rgb]{0,0,0}\makebox(0,0)[lb]{\smash{\textbf{$W(C)$}}}}%
    \put(0.17396213,0.02300807){\color[rgb]{0,0,0}\makebox(0,0)[lb]{\smash{\textbf{(d)}}}}%
    \put(0,0){\includegraphics[width=\unitlength,page=8]{WOn1loopg4.pdf}}%
    \put(0.18367293,0.08951687){\color[rgb]{0,0,0}\makebox(0,0)[lb]{\smash{$1$}}}%
    \put(0.21805931,0.19626882){\color[rgb]{0,0,0}\makebox(0,0)[lb]{\smash{$O_{\vec n}(x)$}}}%
    \put(0,0){\includegraphics[width=\unitlength,page=9]{WOn1loopg4.pdf}}%
    \put(0.7023539,0.26085401){\color[rgb]{0,0,0}\makebox(0,0)[lb]{\smash{\textbf{$W(C)$}}}}%
    \put(0.8463539,0.26085401){\color[rgb]{0,0,0}\makebox(0,0)[lb]{\smash{\textbf{(c)}}}}%
    \put(0,0){\includegraphics[width=\unitlength,page=10]{WOn1loopg4.pdf}}%
    \put(0.84116498,0.3553222){\color[rgb]{0,0,0}\makebox(0,0)[lb]{\smash{$1$}}}%
    \put(0.88834158,0.43428981){\color[rgb]{0,0,0}\makebox(0,0)[lb]{\smash{$O_{\vec n}(x)$}}}%
    \put(0,0){\includegraphics[width=\unitlength,page=11]{WOn1loopg4.pdf}}%
    \put(0.55208137,0.19626882){\color[rgb]{0,0,0}\makebox(0,0)[lb]{\smash{$O_{\vec n}(x)$}}}%
    \put(0.3695401,0.02300807){\color[rgb]{0,0,0}\makebox(0,0)[lb]{\smash{\textbf{$W(C)$}}}}%
    \put(0.5145401,0.02300807){\color[rgb]{0,0,0}\makebox(0,0)[lb]{\smash{\textbf{(e)}}}}%
    \put(0,0){\includegraphics[width=\unitlength,page=12]{WOn1loopg4.pdf}}%
    \put(0.88888132,0.19626882){\color[rgb]{0,0,0}\makebox(0,0)[lb]{\smash{$O_{\vec n}(x)$}}}%
    \put(0.70634009,0.02300807){\color[rgb]{0,0,0}\makebox(0,0)[lb]{\smash{\textbf{$W(C)$}}}}%
    \put(0.85134009,0.02300807){\color[rgb]{0,0,0}\makebox(0,0)[lb]{\smash{\textbf{(f)}}}}%
    \put(0,0){\includegraphics[width=\unitlength,page=13]{WOn1loopg4.pdf}}%
    \put(0.8492722,0.07644654){\color[rgb]{0,0,0}\makebox(0,0)[lb]{\smash{$1$}}}%
    \put(0,0){\includegraphics[width=\unitlength,page=14]{WOn1loopg4.pdf}}%
    \put(0.89612413,0.13644283){\color[rgb]{0,0,0}\makebox(0,0)[lb]{\smash{$1$}}}%
    \put(0,0){\includegraphics[width=\unitlength,page=15]{WOn1loopg4.pdf}}%
    \put(0.49035591,0.16171697){\color[rgb]{0,0,0}\makebox(0,0)[lb]{\smash{$1$}}}%
    \put(0,0){\includegraphics[width=\unitlength,page=16]{WOn1loopg4.pdf}}%
    \put(0.53606163,0.10659241){\color[rgb]{0,0,0}\makebox(0,0)[lb]{\smash{$1$}}}%
  \end{picture}%
\endgroup%
\end{center}
\caption{Some examples of diagrams contributing to $\delta A_{\vec{n}}$ at two loops. 
Diagrams $\mathbf{(a)}$ and $\mathbf{(c)}$ contain the one-loop correction of the 
gluon propagator, diagram $\mathbf{(d)}$ contains the one-loop 
correction to the 3-point vertex, while diagrams 
$\mathbf{(b)}$, $\mathbf{(e)}$ and $\mathbf{(f)}$ contain the one-loop correction to the
scalar propagator. All these diagrams vanish in the superconformal theory since they are
proportional to $(N_f-2N)$. Beside these, one should also consider the one-loop 
diagrams of Fig.~\ref{fig:WOnN4} in which one of the external scalar propagators is corrected 
at one loop. Also such diagrams vanish in the superconformal theory.}
\label{fig:WOn1loopg4}
\end{figure}

Let us now consider the two-loop corrections, {\it{i.e.}} those at order $g^4$ with respect to the
tree-level amplitudes, and focus on the difference $\delta A_{\vec{n}}$.
The $H$ or $Q$ diagrams which contribute at this order can be divided into two classes. The first one
is formed by those diagrams which contain a sub-diagram with the one-loop correction to the scalar
propagator, or to the gluon propagator or to the 3-point vertex. Some examples of such diagrams are shown in Fig.~\ref{fig:WOn1loopg4}.
All these diagrams vanish in the $\cN=2$ superconformal theory. 
Indeed, both the one-loop correction to the 
gluon propagator and the one-loop correction to the 3-point vertex are 
proportional to $(N_f-2N)$, just like
the one-loop correction to the scalar propagator as we have seen in (\ref{color1loop})

\begin{figure}[ht]
\vspace{0.3cm}
\begin{center}
\begingroup%
  \makeatletter%
  \providecommand\color[2][]{%
    \errmessage{(Inkscape) Color is used for the text in Inkscape, but the package 'color.sty' is not loaded}%
    \renewcommand\color[2][]{}%
  }%
  \providecommand\transparent[1]{%
    \errmessage{(Inkscape) Transparency is used (non-zero) for the text in Inkscape, but the package 'transparent.sty' is not loaded}%
    \renewcommand\transparent[1]{}%
  }%
  \providecommand\rotatebox[2]{#2}%
  \ifx\svgwidth\undefined%
    \setlength{\unitlength}{305bp}%
    \ifx\svgscale\undefined%
      \relax%
    \else%
      \setlength{\unitlength}{\unitlength * \real{\svgscale}}%
    \fi%
  \else%
    \setlength{\unitlength}{\svgwidth}%
  \fi%
  \global\let\svgwidth\undefined%
  \global\let\svgscale\undefined%
  \makeatother%
  \begin{picture}(1,0.31558777)%
    \put(0,0){\includegraphics[width=\unitlength,page=1]{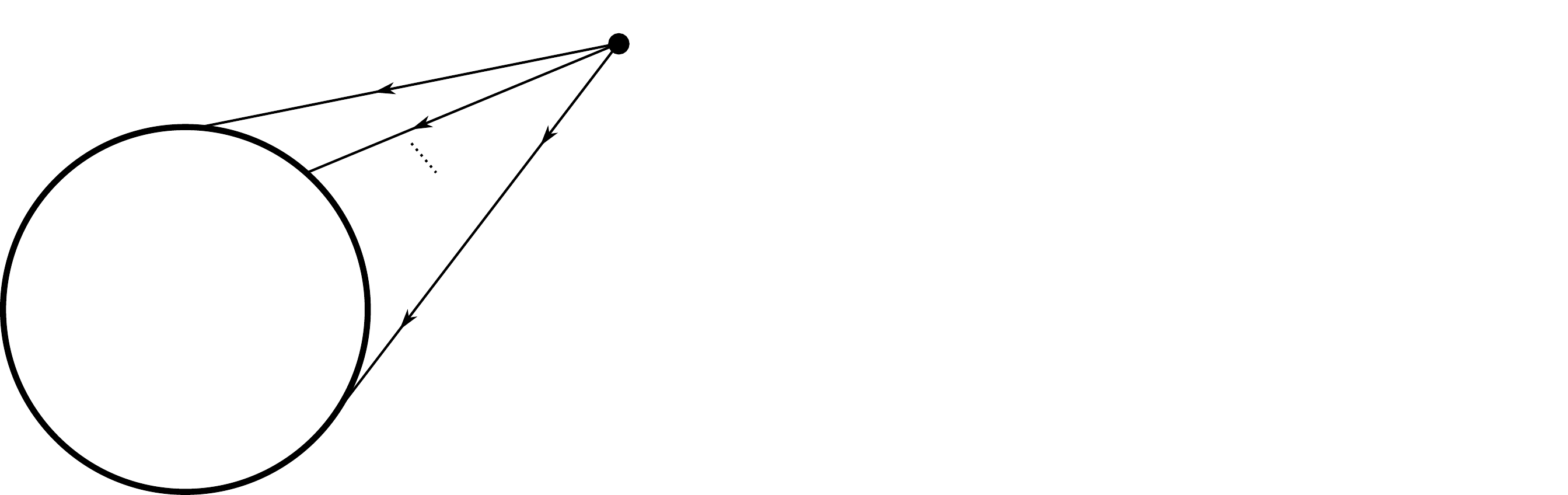}}%
    \put(0.33890409,0.30380454){\color[rgb]{0,0,0}\makebox(0,0)[lb]{\smash{$O_{\vec n}(x)$}}}%
    \put(0.05647463,0.04107561){\color[rgb]{0,0,0}\makebox(0,0)[lb]{\smash{$W(C)$}}}%
\put(0.27647463,0.04107561){\color[rgb]{0,0,0}\makebox(0,0)[lb]{\smash{\textbf{(i)}}}}%
    \put(0,0){\includegraphics[width=\unitlength,page=2]{WOn2loop.pdf}}%
    \put(0.29708383,0.16184158){\color[rgb]{0,0,0}\makebox(0,0)[lb]{\smash{$2$}}}%
    \put(0,0){\includegraphics[width=\unitlength,page=3]{WOn2loop.pdf}}%
    \put(0.83325455,0.30380454){\color[rgb]{0,0,0}\makebox(0,0)[lb]{\smash{$O_{\vec n}(x)$}}}%
    \put(0.56082511,0.04107561){\color[rgb]{0,0,0}\makebox(0,0)[lb]{\smash{$W(C)$}}}%
    \put(0,0){\includegraphics[width=\unitlength,page=4]{WOn2loop.pdf}}%
    \put(0.73047399,0.23086561){\color[rgb]{0,0,0}\makebox(0,0)[lb]{\smash{$2$}}}%
\put(0.78582511,0.04107561){\color[rgb]{0,0,0}\makebox(0,0)[lb]{\smash{\textbf{(j)}}}}%
  \end{picture}%
\endgroup%
\end{center}
\caption{Diagrams that contribute to $\delta A_{\vec{n}}$ at two loops in the $\cN=2$ superconformal
theory. 
Diagram $\mathbf{(i)}$ on the left contains the irreducible two-loop correction of the scalar propagator
represented in Fig.~\ref{fig:scalprop2l}, while diagram $\mathbf{(j)}$ on the right contains the 
two-loop effective vertex depicted in Fig.~\ref{fig:scalbox2l}.}
\label{fig:WOn2loop}
\end{figure}

\begin{figure}[ht]
\vspace{0.3cm}
\begin{center}
\begingroup%
  \makeatletter%
  \providecommand\color[2][]{%
    \errmessage{(Inkscape) Color is used for the text in Inkscape, but the package 'color.sty' is not loaded}%
    \renewcommand\color[2][]{}%
  }%
  \providecommand\transparent[1]{%
    \errmessage{(Inkscape) Transparency is used (non-zero) for the text in Inkscape, but the package 'transparent.sty' is not loaded}%
    \renewcommand\transparent[1]{}%
  }%
  \providecommand\rotatebox[2]{#2}%
  \ifx\svgwidth\undefined%
    \setlength{\unitlength}{300bp}%
    \ifx\svgscale\undefined%
      \relax%
    \else%
      \setlength{\unitlength}{\unitlength * \real{\svgscale}}%
    \fi%
  \else%
    \setlength{\unitlength}{\svgwidth}%
  \fi%
  \global\let\svgwidth\undefined%
  \global\let\svgscale\undefined%
  \makeatother%
  \begin{picture}(1,0.28113794)%
    \put(0,0){\includegraphics[width=\unitlength,page=1]{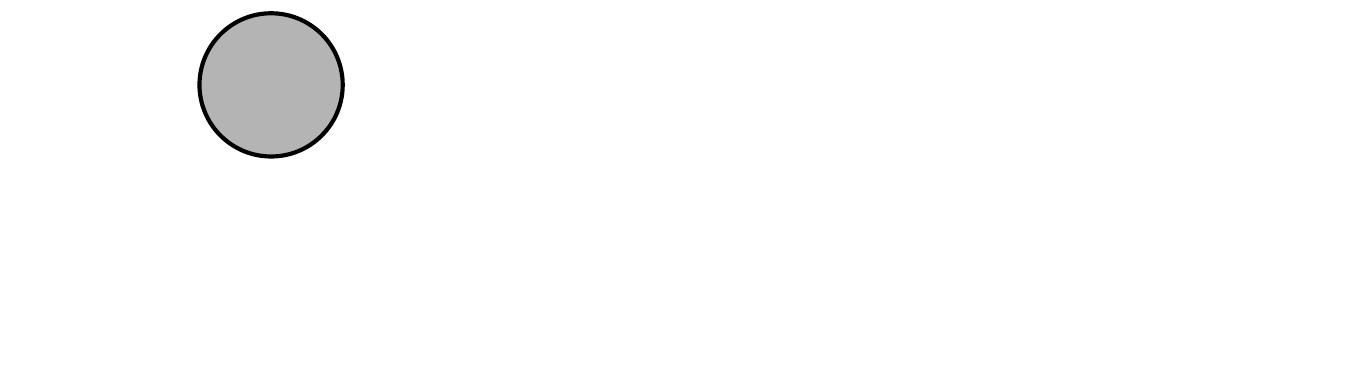}}%
    \put(0.18945397,0.21098425){\color[rgb]{0,0,0}\makebox(0,0)[lb]{\smash{$2$}}}%
    \put(0,0){\includegraphics[width=\unitlength,page=2]{scalprop2l.pdf}}%
    \put(0.4506696,0.20827093){\color[rgb]{0,0,0}\makebox(0,0)[lb]{\smash{\textbf{$=$}}}}%
    \put(0,0){\includegraphics[width=\unitlength,page=3]{scalprop2l.pdf}}%
    \put(0.49331781,0.05252796){\color[rgb]{0,0,0}\makebox(0,0)[lb]{\smash{\textbf{$-$}}}}%
    \put(0,0){\includegraphics[width=\unitlength,page=4]{scalprop2l.pdf}}%
    \put(-0.00157482,0.24010037){\color[rgb]{0,0,0}\makebox(0,0)[lb]{\smash{\textbf{$b$}}}}%
    \put(0.38306589,0.24010037){\color[rgb]{0,0,0}\makebox(0,0)[lb]{\smash{\textbf{$a$}}}}%
    \put(0.57425651,0.24010037){\color[rgb]{0,0,0}\makebox(0,0)[lb]{\smash{\textbf{$b$}}}}%
    \put(0.95011344,0.24010037){\color[rgb]{0,0,0}\makebox(0,0)[lb]{\smash{\textbf{$a$}}}}%
    \put(0.57425651,0.08102176){\color[rgb]{0,0,0}\makebox(0,0)[lb]{\smash{\textbf{$b$}}}}%
    \put(0.95011344,0.08102176){\color[rgb]{0,0,0}\makebox(0,0)[lb]{\smash{\textbf{$a$}}}}%
  \end{picture}%
\endgroup%
\end{center}
\caption{The irreducible two-loop correction to the scalar propagator in the $\cN=2$ superconformal theory.
The first diagram on the right hand side is the $Q$-contribution involving the fundamental hypermultiplets, while the second diagram is the $H$-contribution due to the adjoint hypermultiplet which therefore comes
with a minus sign.}
\label{fig:scalprop2l}
\end{figure}

\begin{figure}[ht]
\vspace{0.3cm}
\begin{center}
\begingroup%
  \makeatletter%
  \providecommand\color[2][]{%
    \errmessage{(Inkscape) Color is used for the text in Inkscape, but the package 'color.sty' is not loaded}%
    \renewcommand\color[2][]{}%
  }%
  \providecommand\transparent[1]{%
    \errmessage{(Inkscape) Transparency is used (non-zero) for the text in Inkscape, but the package 'transparent.sty' is not loaded}%
    \renewcommand\transparent[1]{}%
  }%
  \providecommand\rotatebox[2]{#2}%
  \ifx\svgwidth\undefined%
    \setlength{\unitlength}{300bp}%
    \ifx\svgscale\undefined%
      \relax%
    \else%
      \setlength{\unitlength}{\unitlength * \real{\svgscale}}%
    \fi%
  \else%
    \setlength{\unitlength}{\svgwidth}%
  \fi%
  \global\let\svgwidth\undefined%
  \global\let\svgscale\undefined%
  \makeatother%
  \begin{picture}(1,0.3090492)%
    \put(0,0){\includegraphics[width=\unitlength,page=1]{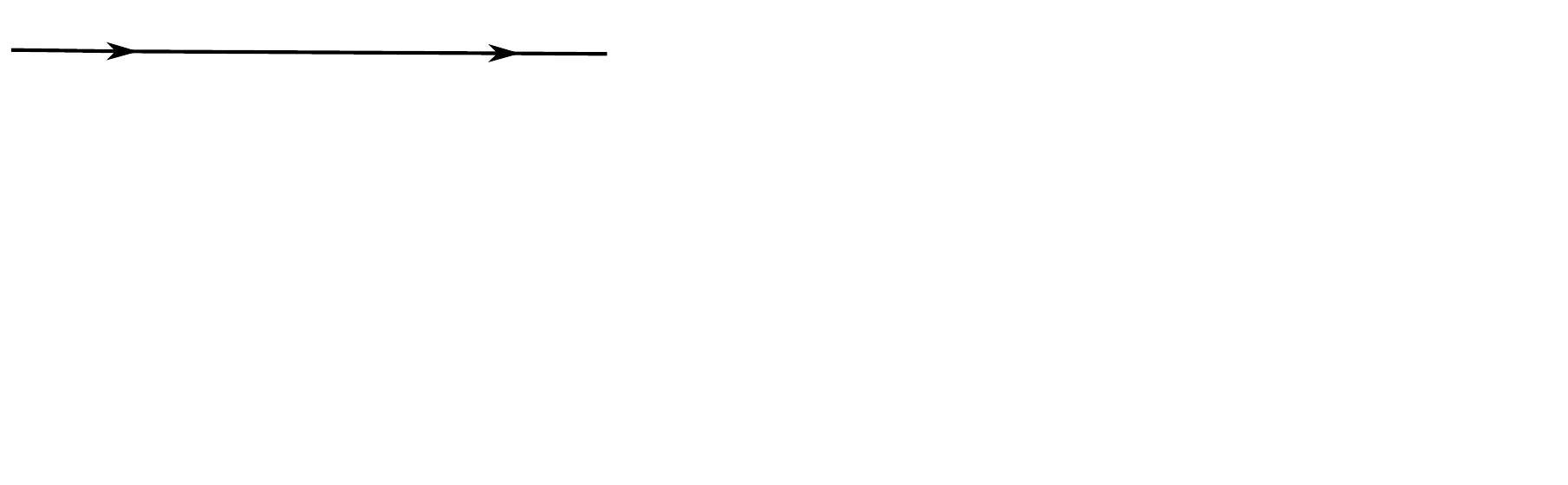}}%
    \put(-0.00070073,0.29117511){\color[rgb]{0,0,0}\makebox(0,0)[lb]{\smash{\textbf{$b_1$}}}}%
    \put(0.36659274,0.29117511){\color[rgb]{0,0,0}\makebox(0,0)[lb]{\smash{\textbf{$a_1$}}}}%
    \put(0,0){\includegraphics[width=\unitlength,page=2]{scalbox2l.pdf}}%
    \put(-0.00117471,0.20750945){\color[rgb]{0,0,0}\makebox(0,0)[lb]{\smash{\textbf{$b_2$}}}}%
    \put(0.36611876,0.20750945){\color[rgb]{0,0,0}\makebox(0,0)[lb]{\smash{\textbf{$a_2$}}}}%
    \put(0.39448535,0.2154346){\color[rgb]{0,0,0}\makebox(0,0)[lb]{\smash{}}}%
    \put(0,0){\includegraphics[width=\unitlength,page=3]{scalbox2l.pdf}}%
    \put(0.17813853,0.22961798){\color[rgb]{0,0,0}\makebox(0,0)[lb]{\smash{$2$}}}%
    \put(0.32538448,0.15401161){\color[rgb]{0,0,0}\makebox(0,0)[lb]{\smash{}}}%
    \put(0.4448685,0.22920113){\color[rgb]{0,0,0}\makebox(0,0)[lb]{\smash{\textbf{$=$}}}}%
    \put(0,0){\includegraphics[width=\unitlength,page=4]{scalbox2l.pdf}}%
    \put(0.55560092,0.29117511){\color[rgb]{0,0,0}\makebox(0,0)[lb]{\smash{\textbf{$b_1$}}}}%
    \put(0.93817956,0.29117511){\color[rgb]{0,0,0}\makebox(0,0)[lb]{\smash{\textbf{$a_1$}}}}%
    \put(0.55560092,0.20750945){\color[rgb]{0,0,0}\makebox(0,0)[lb]{\smash{\textbf{$b_2$}}}}%
    \put(0.93817956,0.20750945){\color[rgb]{0,0,0}\makebox(0,0)[lb]{\smash{\textbf{$a_2$}}}}%
    \put(0,0){\includegraphics[width=\unitlength,page=5]{scalbox2l.pdf}}%
    \put(0.4848685,0.04355529){\color[rgb]{0,0,0}\makebox(0,0)[lb]{\smash{\textbf{$-$}}}}%
    \put(0,0){\includegraphics[width=\unitlength,page=6]{scalbox2l.pdf}}%
    \put(0.55560092,0.10552926){\color[rgb]{0,0,0}\makebox(0,0)[lb]{\smash{\textbf{$b_1$}}}}%
    \put(0.93817956,0.10552926){\color[rgb]{0,0,0}\makebox(0,0)[lb]{\smash{\textbf{$a_1$}}}}%
    \put(0.55560092,0.02186361){\color[rgb]{0,0,0}\makebox(0,0)[lb]{\smash{\textbf{$b_2$}}}}%
    \put(0.93817956,0.02186361){\color[rgb]{0,0,0}\makebox(0,0)[lb]{\smash{\textbf{$a_2$}}}}%
    \put(0,0){\includegraphics[width=\unitlength,page=7]{scalbox2l.pdf}}%
  \end{picture}%
\endgroup%
\end{center}
\caption{The two-loop effective vertex that can contribute to the amplitude $A_{\vec{n}}$
in the $\cN=2$ superconformal theory.
The first diagram on the right hand side is the $Q$-contribution involving the fundamental hypermultiplets, while the second diagram is the $H$-contribution due to the adjoint hypermultiplet and thus comes
with a minus sign.}
\label{fig:scalbox2l}
\end{figure}

The second class of diagrams that can contribute to $\delta A_{\vec{n}}$ at two loops in the
superconformal theory are those of the type displayed in Fig.~\ref{fig:WOn2loop}.
They contain either the irreducible two-loop correction of the scalar propagator 
represented in Fig.~\ref{fig:scalprop2l}, or the two-loop effective vertex represented 
in Fig.~\ref{fig:scalbox2l}. Thus, we can write
\begin{equation}
\delta A_{\vec{n}}\Big|_{\mathrm{2-loop}} = \,I_{\vec{n}}+ J_{\vec{n}}
\label{deltaAn2loop}
\end{equation}
where $I_{\vec{n}}$ and $J_{\vec{n}}$ correspond, respectively, to the diagrams of type $\mathbf{(i)}$
and $\mathbf{(j)}$.

Let us first consider the irreducible two-loop correction\,%
\footnote{Notice that in the superconformal theory the only diagrams that contribute to the scalar 
propagator at two loops are those represented in Fig.~\ref{fig:scalprop2l}. Indeed, all other two-loop
diagrams that correct the propagators are proportional to $(N_f-2N)$.} 
of the scalar propagator drawn in Fig.~\ref{fig:scalprop2l}.
In configuration space this correction has been computed in \cite{Billo:2017glv} to which we refer 
for details, and the result is\,%
\footnote{See Eq.~(3.24) of \cite{Billo:2017glv}.}
\begin{equation}
\label{propsub}
-8\, g^4\,C_2^{ba}\, W_2(x_1,x_2)
\end{equation}
where the colour factor is
\begin{equation}
\begin{aligned}
C_2^{ba}&=N_f\,\tr\big(T^bT^cT^aT^c\big)-f^{bd_4d_1}\,f^{cd_1d_2}\,f^{ad_2d_3}\,f^{cd_3d_4}\\
&=-\Big(\frac{N_f}{2N}+N^2\Big) \tr\big(T^bT^a\big)=-\frac{N^2+1}{2}\,\delta^{ab}~,
\end{aligned}
\label{C2}
\end{equation}
while the superspace integral yields
\begin{equation}
W_2(x_1,x_2)=-\frac{3\, \zeta(3)}{(16 \pi^2)^2}\,\frac{1}{4\pi^2(x_1-x_2)^2}~.
\label{W2}
\end{equation}
Putting everything together, we find that the two-loop correction of the scalar propagator is
\begin{equation}
-g^4\,\frac{3\,\zeta(3)}{(8\pi^2)^2}\Big[\frac{\delta^{ab}}{4\pi^2(x_1-x_2)^2}\Big](N^2+1)
\label{scalprop2la}
\end{equation}
where the expression in square brackets is the tree-level propagator. Therefore, when we compute
the amplitude $I_{\vec{n}}$ corresponding to the diagram $\mathbf{(i)}$ of Fig.~\ref{fig:WOn2loop}, 
we simply obtain an expression which is proportional to the tree-level result (\ref{Antree0}). Indeed we 
get
\begin{equation}
I_{\vec{n}}= -n\,g^4\,\frac{3\,\zeta(3)}{(8\pi^2)^2}\Big[\frac{g^n}{N\,2^{\frac{n}{2}}}
\,R_{\vec{n}}^{\,b_1\dots b_n}\,\tr \big(T^{b_1}\dots T^{b_n}\big)\Big](N^2+1)
\label{In}
\end{equation}
where the overall factor of $n$ is due to the fact that the two-loop correction (\ref{scalprop2la})
can be inserted in any of the $n$ external propagators.

Let us now consider the two-loop diagram $\mathbf{(j)}$ of Fig.~\ref{fig:WOn2loop}. 
To compute the corresponding amplitude $J_{\vec{n}}$, we have to perform all contractions as in 
the tree-level diagram but with two scalar propagators replaced by the sub-structure corresponding
to the two-loop effective vertex of Fig.~\ref{fig:scalbox2l}. The latter has been analyzed
in \cite{Billo:2017glv} to which we refer again for details. Considering that the two external legs with
colour indices $b_1$ and $b_2$ are inserted at the point $x$ where the operator $O_{\vec{n}}$ is located,
and the other two external legs with indices $a_1$ and $a_2$ are inserted at two points $x_1$ and $x_2$
on the circular Wilson loop, the relevant expression is\,%
\footnote{See Eq.~(3.33) of \cite{Billo:2017glv}.}
\begin{equation}
2\,g^4\,C_4^{\,b_1b_2a_1a_2}\,W_4(x,x;x_1,x_2)
\label{boxsub}
\end{equation}
where the colour factor is
\begin{equation}
\begin{aligned}
C_4^{\,b_1b_2a_1a_2}&=N_f\,\tr\big(T^{b_1}T^{a_1}T^{b_2}T^{a_2}\big)
-f^{b_1d_4d_1}\,f^{a_1d_1d_2}\,f^{b_2d_2d_3}\,f^{a_2d_3d_4}\\
&=-\frac{1}{2}\big(\delta^{b_1a_1}\,\delta^{b_2a_2}+\delta^{b_1 b_2}\,\delta^{a_1 a_2}
+\delta^{b_1a_2}\,\delta^{b_2a_1}\big)~,
\label{C4}
\end{aligned}
\end{equation}
while the superspace integral leads to
\begin{equation}
W_4(x,x;x_1,x_2)=\frac{6\, \zeta(3)}{(16 \pi^2)^2}\,\Big[
\frac{1}{4\pi^2(x-x_{1})^2}\,\frac{1}{4\pi^2(x-x_{2})^2}\Big]~.
\label{W4}
\end{equation}
As is clear from the expression in square brackets, we still recover the same space dependence of two 
scalar propagators as in the tree-level computation, even if the colour structure of the $C_4$ tensor is
different. Exploiting conformal invariance to set $x=0$ and recalling the parametrization (\ref{circle})
for points on a circle, the above square brackets simply becomes $1/(2\pi R)^4$;
thus the path-ordering and the integration over the Wilson loop become trivial to perform, just
as they were in the tree-level amplitude. Putting everything together and replacing any pair of external
scalar propagators with this effective two-loop vertex in all possible ways, we obtain
\begin{equation}
\begin{aligned}
J_{\vec{n}}&=g^4\,\frac{3\,\zeta(3)}{(8\pi^2)^2}\,
\Big[\frac{g^n}{N\,2^{\frac{n}{2}}}\,R_{\vec{n}}^{\,b_1\dots b_n}\,
\tr \big(T^{a_1}\dots T^{a_n}\big)\Big]
\\
&~~~\times 2\!\!\!
\sum_{p \in S_{n-1}} \!\!\!C_4^{\,b_1b_2 a_{p(1)} a_{p(2)}}
\, \delta^{b_3a_{p(3)}}
\dots\delta^{b_{n-1}a_{p(n-1)}}\,\delta^{b_n a_n}
\end{aligned}
\label{Jn}
\end{equation}
where $p\in S_{n-1}$ are the permutations of $(n-1)$ elements.
We observe that the $1/n!$ coming from the expansion of the Wilson loop operator at order $g^n$
is compensated by a factor of $n!$ that arises when we take into account the complete symmetry
of the tensor $R_{\vec{n}}$ and the cyclic symmetry of the trace factor in the square bracket. 
Furthermore the factor of 2 in the last line of (\ref{Jn}) is a combinatorial factor due to the 
multiplicity of the two-loop box diagram of Fig.~\ref{fig:scalbox2l}.

Summing $I_{\vec{n}}$ and $J_{\vec{n}}$, we get
\begin{align}
\delta A_{\vec{n}}\Big|_{\mathrm{2-loop}}&= -g^4\,\frac{3\, \zeta(3)}{(8 \pi^2)^2}\,
\Big[\frac{g^n}{N\,2^{\frac{n}{2}}}\,R_{\vec{n}}^{\,b_1\dots b_n}\,
\tr \big(T^{a_1}\dots T^{a_n}\big)\Big]\phantom{\bigg|}\label{deltaAn2loopfin}\\
&\hspace{-0.75cm}
\times \Big[n\,(N^2+1)\,\delta^{b_1a_1}\dots\delta^{b_na_n}-2\!\!\!
\sum_{p \in S_{n-1}} \!\!\!C_4^{\,b_1b_2 a_{p(1)} a_{p(2)}}
\, \delta^{b_3a_{p(3)}}
\dots\delta^{b_{n-1}a_{p(n-1)}}\,\delta^{b_n a_n}\,\Big]~.\phantom{\bigg|}\notag
\end{align}
This is the final result of our diagrammatic computation of the two-loop correction to the
amplitude $A_{\vec{n}}$ in the $\cN=2$ superconformal theory.

As an example, we work out the explicit expression for the lowest dimensional operator $O_{(2)}$. In this
case, we simply have
\begin{equation}
R_{(2)}^{b_1b_2}=\tr\big(T^{b_1}T^{b_2}\big)=\frac{1}{2}\,\delta^{b_1b_2}~.
\end{equation}
Thus, the contribution from the diagram $\mathbf{(i)}$ is (see (\ref{In})):
\begin{equation}
I_{(2)}= -2\,g^4\,\frac{3\,\zeta(3)}{(8\pi^2)^2}\Big[\frac{g^2}{2\,N}
\frac{(N^2-1)}{4}\Big](N^2+1)~,
\label{I2}
\end{equation}
while from the diagram $\mathbf{(j)}$ we get (see (\ref{Jn})):
\begin{equation}
J_{(2)}= -g^4\,\frac{3\,\zeta(3)}{(8\pi^2)^2}\Big[\frac{g^2}{2\,N}\frac{(N^2-1)}{4}\Big]
(N^2+1)~.\label{J2}
\end{equation}
Note that in this case both diagrams $\mathbf{(i)}$ and $\mathbf{(j)}$ provide colour contributions with the same leading power of $N$. This is a specific feature of this operator and it does not hold for higher dimensional operators unless they contain a factor of $\tr \phi^2$ (see Appendix~\ref{app:color}
where we discuss the cases corresponding to $\vec{n}=(4)$ and $\vec{n}=(2,2)$ in which this property
is clearly exhibited). This fact will have important consequences 
for the planar limit as we will see in the following subsection.
Summing (\ref{I2}) and (\ref{J2}), we finally get
\begin{equation}
\delta A_{(2)}\Big|_{\mathrm{2-loop}}= -g^6\,\frac{\zeta(3)}{(8\pi^2)^2}\,
\frac{9(N^2-1)(N^2+1)}{8N}~,
\label{deltaA22loopft}
\end{equation}
in perfect agreement with the matrix model result (\ref{deltaA22loop}).

We have explicitly performed similar checks for many operators of higher dimension and always 
found a precise match between the field theory expression (\ref{deltaAn2loopfin}) and the matrix 
model results summarized in Tab.~\ref{tab2}, thus confirming the validity of (\ref{checkN2N4})
up to two loops. The details of the calculation in the cases $\vec{n}=(4)$ and $\vec{n}=(2,2)$
are given in Appendix~\ref{app:color}.

\subsection{Planar limit}
All the above checks are easily extended in the planar limit by keeping the highest power of $N$ and performing the substitution $g^2N=\lambda$. In this limit the number of diagrams which contribute to the correlator is drastically reduced, and thus such checks can be pushed to higher orders in perturbation theory without much effort. Let us first review the well-known
$\cN=4$ case \cite{Semenoff:2001xp,Pestun:2002mr,Semenoff:2006am}.

\subsubsection*{The $\cN=4$ theory}
At leading order, using the tree-level result (\ref{Antree0}) that corresponds to the diagram
of Fig.~\ref{fig:WOntree}, one easily finds
\begin{equation}
g^{n-2\ell}\,\widehat A_{\vec n}\Big|_{\mathrm{tree-level,planar}} 
=\lim_{N\to \infty}
\frac{g^{2n-2\ell}}{N\,2^{\frac{n}{2}}}\,R_{\vec{n}}^{\,b_1\dots b_n}\,\tr \big(T^{b_1}\dots T^{b_n}\big)
=c_{\vec{n},0}\,\lambda^{n-\ell}
\end{equation}
where $c_{\vec{n},0}$ are numerical coefficients which can be deduced from Tab.~\ref{tab1}. In particular
we have:
\begin{equation}
c_{(2),0}=\frac{1}{8}~,~~c_{(3),0}=\frac{1}{32\sqrt{2}}~,~~c_{(4),0}=\frac{1}{384}
~,~~c_{(2,2),0}=\frac{1}{96}~.
\end{equation}

\begin{figure}[ht]
\begin{center}
\begingroup%
  \makeatletter%
  \providecommand\color[2][]{%
    \errmessage{(Inkscape) Color is used for the text in Inkscape, but the package 'color.sty' is not loaded}%
    \renewcommand\color[2][]{}%
  }%
  \providecommand\transparent[1]{%
    \errmessage{(Inkscape) Transparency is used (non-zero) for the text in Inkscape, but the package 'transparent.sty' is not loaded}%
    \renewcommand\transparent[1]{}%
  }%
  \providecommand\rotatebox[2]{#2}%
  \ifx\svgwidth\undefined%
    \setlength{\unitlength}{320bp}%
    \ifx\svgscale\undefined%
      \relax%
    \else%
      \setlength{\unitlength}{\unitlength * \real{\svgscale}}%
    \fi%
  \else%
    \setlength{\unitlength}{\svgwidth}%
  \fi%
  \global\let\svgwidth\undefined%
  \global\let\svgscale\undefined%
  \makeatother%
  \begin{picture}(1,0.31167873)%
    \put(0,0){\includegraphics[width=\unitlength,page=1]{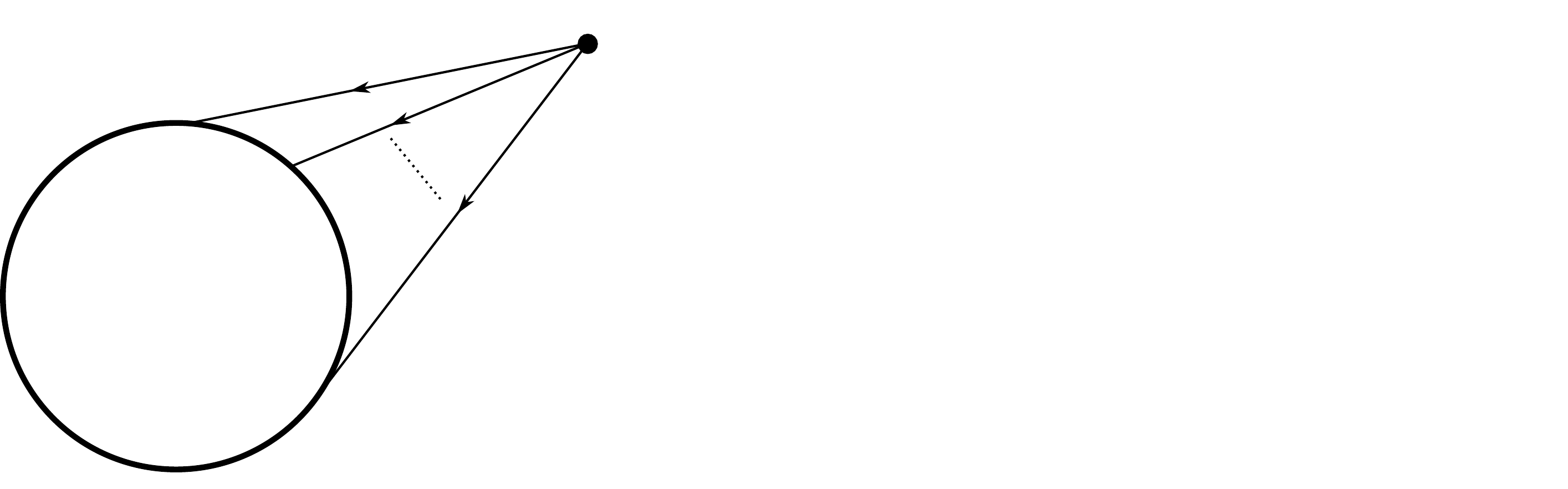}}%
    \put(0.30995034,0.29904349){\color[rgb]{0,0,0}\makebox(0,0)[lb]{\smash{$O_{\vec n}(x)$}}}%
    \put(0.04147309,0.04158717){\color[rgb]{0,0,0}\makebox(0,0)[lb]{\smash{\textbf{$W(C)$}}}}%
    \put(0.27947309,0.04158717){\color[rgb]{0,0,0}\makebox(0,0)[lb]{\smash{\textbf{(r)}}}}%
    \put(0,0){\includegraphics[width=\unitlength,page=2]{WOnN4rain.pdf}}%
    \put(0.55537636,0.04158717){\color[rgb]{0,0,0}\makebox(0,0)[lb]{\smash{\textbf{$W(C)$}}}}%
    \put(0.79337636,0.04158717){\color[rgb]{0,0,0}\makebox(0,0)[lb]{\smash{\textbf{(s)}}}}%
    \put(0.82959469,0.30048493){\color[rgb]{0,0,0}\makebox(0,0)[lb]{\smash{$O_{\vec n}(x)$}}}%
    \put(0,0){\includegraphics[width=\unitlength,page=3]{WOnN4rain.pdf}}%
  \end{picture}%
\endgroup%
\end{center}
\caption{In the planar limit of the $\cN=4$ theory, the tree-level expression encoded in 
Fig.~\ref{fig:WOntree} gets corrected only by the so-called ``rainbow'' diagrams, the first two of which are
represented here. We have used a double line to denote the sum of the gluon and the scalar propagator, which always occur together when attached to the Wilson loop and yield the simple expression given in (\ref{ww}).}
\label{fig:WOnN4rain}
\end{figure}

In \cite{Semenoff:2001xp} it was argued that all diagrams with internal vertices cancel at the next order 
and it was conjectured that analogous cancellations should occur at all orders in perturbation theory. Thus,
only the ``rainbow'' diagrams of the type represented in Fig.~\ref{fig:WOnN4rain} contribute to the
amplitude $\widehat{A}_{\vec{n}}$ in the planar limit.

The evaluation of these ``rainbow'' diagrams is particularly simple in the case of a circular Wilson loop. Indeed, 
if we denote by $w^a(x)$ the combination of gluons and scalars that appears inside the 
path-ordered exponential in (\ref{WLdef2}), namely
\begin{equation}
w^a(x)=
\ii \,A^a_{\mu}(x)\,\dot{x}^{\mu}
+\frac{R}{\sqrt{2}}\Big(\varphi^a(x) + \bar\varphi^a(x)\Big)
\end{equation}
with $x$ being a point on the circle $C$, then we have
\begin{equation}
\big\langle
w^a(x_1)\,w^b(x_2)\big\rangle=\frac{\delta^{ab}}{4\pi^2}\,\frac{1-\dot{x_1}\cdot\dot{x_2}}{(x_1
-x_2)^2}=\frac{\delta^{ab}}{8\pi^2R^2}
\label{ww}
\end{equation}
where in the last step we have used the parameterization (\ref{circle}). Thus, the contribution of the
internal propagators, represented by double lines in Fig.~\ref{fig:WOnN4rain}, is constant and 
similar to the one of the external scalar propagators (see (\ref{scalprop})) so that 
only combinatorial coefficients have to be computed. For example, the diagram $\mathbf{(r)}$ yields a contribution of the form
\begin{equation}
c_{\vec{n},1}\,\lambda^{n-\ell+1}
\end{equation}
with
\begin{equation}
c_{(2),1}=\frac{1}{96}~,~~c_{(3),1}=\frac{1}{512\sqrt{2}}~,~~c_{(4),1}=\frac{1}{7680}
~,~~c_{(2,2),1}=\frac{1}{1536}~.
\end{equation}
Similarly, the diagram $\mathbf{(s)}$ leads to
\begin{equation}
c_{\vec{n},2}\,\lambda^{n-\ell+2}
\end{equation}
with
\begin{equation}
c_{(2),2}=\frac{1}{3072}~,~~c_{(3),2}=\frac{1}{24480\sqrt{2}}~,~~c_{(4),2}=\frac{1}{368640}
~,~~c_{(2,2),2}=\frac{1}{61440}~.
\end{equation}
{From} these results it is possible to infer the following resummed expression
\begin{equation}
g^{n-2\ell}\,\widehat{A}_{\vec{n}}\Big|_{\mathrm{planar}} =
\sum_{j=0}^\infty c_{\vec{n},j}\,\lambda^{n-\ell+j}
= \frac{\big(\sqrt{\lambda}\big)^{n-\ell-1}}{2^{\frac{n}{2}+\ell-1}\phantom{\big|}}
\,I_{n-\ell+1}\big(\sqrt{\lambda}\big)\,\prod_{i=1}^\ell n_i~
\label{hatAnplanarQFT}
\end{equation}
which agrees with the matrix model result (\ref{hatAnplanar}).

\subsubsection*{The $\cN=2$ theory}
In this case we focus on the planar limit of the difference $\delta A_{\vec{n}}$ and in particular
on the terms proportional to $\zeta(3)$. To obtain the result at the lowest order, one simply has to take 
the two-loop result (\ref{deltaAn2loopfin}) and evaluate it in the large-$N$ limit. As we have seen
in the previous subsection, there are two types of terms: one corresponding to the diagram 
$\mathbf{(i)}$ of Fig.~\ref{fig:WOn2loop} and one corresponding to the diagram $\mathbf{(j)}$, which arise from the two-loop contributions 
depicted, respectively, 
in Fig. \ref{fig:scalprop2l} and \ref{fig:scalbox2l}. The correction to the scalar propagator
gives rise to a contribution that always survives in the planar; in fact in (\ref{scalprop2la}) 
it was proved to be proportional to $g^4(N^2+1)$, which in the planar limit reduces to $\lambda^2$.
On the other hand, the two-loop effective vertex does not always contribute in the planar limit, since
it is leading for $N\to \infty$ only when it is attached to $\tr \varphi^2$. This can be realized by noticing that in this case such a diagram, because of (\ref{C4}), always produces the structure
\begin{equation}
\tr\big(T^{b_1}T^{b_2}\big)\delta^{b_1 b_2}\,\delta^{a_1 a_2}
=\frac{1}{2}(N^2-1)\delta^{a_1 a_2}~,
\end{equation}
with the $N^2$ factor making the contribution leading. Thus, the diagrams of type $\mathbf{(i)}$
always contribute in the planar limit, while the diagrams of type $\mathbf{(j)}$ are sub-leading unless
some of the components of the vector $\vec{n}$ are equal to 2.
This fact can be checked in the explicit computations for $O_{(2)}$ (see (\ref{I2}) and (\ref{J2})) 
and for $O_{(4)}$ and $O_{(2,2)}$ reported in Appendix~\ref{app:color}. These simple considerations 
give a nice field theory interpretation to some of the matrix model results presented in Section \ref{secn:largeN}.

Building on the idea that all diagrams with internal vertices cancel at all orders in perturbation theory, like in the $\cN=4$ model \cite{Semenoff:2001xp}, one can construct a class of $\zeta(3)$-proportional diagrams, starting from the $\cN=4$ ``rainbow'' diagrams and performing on them one of the aforementioned 
planar two-loop corrections. This can be done either by correcting one of the external scalar propagators, 
or by correcting one of the internal double-line propagators\,\footnote{Since these internal 
propagators
and the scalar propagators are proportional to each other (see (\ref{ww}) and (\ref{scalprop})), also their 
planar two-loop corrections are proportional.} or by including the two-loop effective vertex if $O_{\vec{n}}$ contains at least a factor $\tr \varphi^2$.
The result of performing any of these corrections is always equal to the original $\cN=4$ ``rainbow'' 
diagram multiplied by $-\frac{3\,\zeta(3)\,\lambda^2}{(8\pi^2)^2}$. This analysis tells us how to get the
$\cN=2$ correction proportional to $\zeta(3)$ in the planar limit starting from the
$\cN=4$ amplitude. In fact, expanding (\ref{hatAnplanarQFT}) for small $\lambda$, 
the term of order $k$ corresponds to a sum over ``rainbow'' diagrams with $(k-n+\ell)$ 
internal propagators and $n$ external ones. Using the method we just described, any such diagram can be corrected once for every internal propagator, once for every external propagator and once for 
every factor $\tr \varphi^2$ appearing in $O_{\vec{n}}$, giving a total of
\begin{equation}
(k-n+\ell)+n+\sum_{i=1}^\ell \delta_{n_i,2}=k+\ell+\sum_{i=1}^\ell \delta_{n_i,2}
\end{equation}
corrections proportional to $-\frac{3\,\zeta(3)\,\lambda^2}{(8\pi^2)^2}$.
This result precisely matches the matrix model expression (\ref{shiftn}) and suggests that this class of diagrams reconstructs the full $\zeta(3)$-term of the $\cN=2$ correlator at all orders in perturbation theory, just like the ``rainbow'' diagrams make up the full $\cN=4$ correlator.

\section{Conclusions}
\label{secn:concl}
We have verified up to two loops in the $\cN=2$ superconformal theory 
that the one-point amplitude $\cA_{\vec{n}}$ of a chiral operator in presence of a circular Wilson loop computed using the matrix model exactly matches
the amplitude $A_{\vec{n}}$ computed using standard field theory methods with (super) Feynman diagrams. We have also discussed the planar limit of the amplitudes and found a perfect agreement
between the two approaches also in this case.
We have performed our checks in many examples with operators of dimensions up to
$n=7$, even if here we have explicitly reported our results only for the low-dimensional
operators up to $n=4$ for brevity.

We would like to remark that in order to obtain this agreement, an essential ingredient
on the matrix model side is the $g$-dependent normal ordering of the chiral operators introduced in
\cite{Billo:2017glv}. This normal ordering prescription is equivalent to the Gram-Schmidt orthogonalization algorithm discussed in \cite{Gerchkovitz:2016gxx} and later in \cite{Rodriguez-Gomez:2016ijh,Rodriguez-Gomez:2016cem,Baggio:2016skg,Giombi:2018qox} in both $\cN=4$ and $\cN=2$ cases. 
In the $\cN=4$ theory, however, this procedure actually does not introduce any $g$-dependence, while in 
the $\cN=2$ examples considered so far in the literature, the $g$-dependent terms of the normal-ordered operators could not be really tested since they affect only higher-loop subleading terms which have not 
been computed. This is the case, for instance, of the two-point functions of chiral operators investigated in \cite{Gerchkovitz:2016gxx} for the superconformal theory, or in \cite{Billo:2017glv}
for the superconformal theory and for a special class of operators in the non-conformal case. On the contrary, 
for the one-point functions in presence of a Wilson loop that we have studied in this paper, such 
$g$-dependence already shows up at two loops, and thus its crucial role for the
agreement with the field theory results could be tested in our two-loop calculations. 

Several extensions and generalizations are possible. For example, one could compute
the one-point functions of chiral operators in presence of Wilson loops that are more general than the
circular one we have considered and that preserve a smaller amount of supersymmetry. 
Another interesting possibility would be to study the two-point functions in presence of a Wilson loop (as in \cite{Buchbinder:2012vr})
and see what kind of information could be extracted from the matrix model in this case.
An even more challenging development would be to consider non-conformal $\cN=2$ theories \cite{Beccaria:2017rbe}
and check whether also in this case the matrix model can be used to obtain the field theory amplitudes.
As is clear from our discussion in Section~\ref{sec:mm}, there is no obstruction to define and 
compute amplitudes in non-conformal $\cN=2$ theories. One simply
has to take into account the fact that several cancellations do not occur any longer when 
$N_f\not=2N$ and thus more terms have to be considered. On the field theory side, instead, 
one has deal with delicate issues related to the renormalization of the coupling constant, 
of the wave-function and of the composite operators, and also to the appearance of a dynamically generated scale at the quantum level. We believe that making some progress in this direction would be
very interesting since the matrix model approach is technically much more amenable than the 
diagrammatic one and allows one to obtain results at high perturbative orders in a more efficient way.

\vskip 1.5cm
\noindent {\large {\bf Acknowledgments}}
\vskip 0.2cm
We thank M. Frau, F. Fucito, R.~R. John, G. Korchemski, E. Lauria, L. Magnea and J.~F. Morales for many useful discussions. 

\noindent
The work of M.B.~and A.L. is partially supported by the MIUR PRIN Contract 
2015 MP2CX4 ``Non-perturbative Aspects Of Gauge Theories And Strings''.
\vskip 1cm
\begin{appendix}
\section{One-point functions from defect conformal field theory}
\label{secn:appa}
In a conformal field theory, the functional form of the one-point function of a conformal operator 
$O(x)$ in presence of a circular defect $W(C)$ of radius $R$ is completely determined. 
One way to obtain this form is to use the embedding formalism, 
in which a point $x\in\mathbb{R}^4$ is associated in a projective way to a null section $P$ in 
the embedding space $\mathbb{M}^{1,5}$ of the form
\begin{equation}
\label{Pdef}
P = \Big(\frac{R^2+x^2}{2R},\frac{R^2-x^2}{2R},x^\mu\Big)~,
\end{equation} 
which satisfies $P^2 \equiv P^T\eta \,P = 0$ with $\eta=\mathrm{diag}(-1,1,1,1,1,1)$.
Scalar operators $O(x)$ of dimension $\Delta$ are associated to
operators $\hat O(P)$ which are homogeneous of degree $\Delta$, namely such that 
$\widehat O(\lambda P) = \lambda^{-\Delta}\, \widehat O(P)$. 
 
In absence of defects, the conformal group SO$(1,5)$ is the isometry group of the embedding space and acts linearly on $P$. In presence of the Wilson loop, we can split the spacetime coordinates into ``parallel'' and ``transverse'' ones: 
$x^\mu \to (x^a,x^i)$, where $a=1,2$ and $i=3,4$. We will denote $x^a x_a = r^2$ and $x^i x_i = L^2$, so that $x^2 = r^2 + L^2$.
The symmetry is reduced according to the pattern
\begin{equation}
\label{sbp}
\mathrm{SO}(1,5)\to \mathrm{SO}(1,2)\times \mathrm{SO}(3)~,
\end{equation} 
with SO$(1,2)$ and SO$(3)$ linearly acting, respectively, on 
\begin{equation}
\label{SOs}
P_\parallel = \Big(\frac{R^2+x^2}{2R},x^a\Big)~~~\text{and}~~~
P_\perp = \Big(\frac{R^2-x^2}{2R},x^i\Big)~.
\end{equation} 
There are two scalar products invariant with respect to the two symmetry factors, 
which  we denote as
\begin{equation}
\label{scpr}
P\gbullet P \equiv P_\parallel^T\eta \,P_\parallel~~\text{with}~
\eta=\mathrm{diag}(-1,1,1)~~~~~\mbox{and}~~~~~
P\wbullet P \equiv P_\perp^T \, P_\perp~. 
\end{equation}
They are not independent, since $P\gbullet P + P\wbullet P = P^2 = 0$. Therefore, we
can take as the single independent invariant the quantity
\begin{equation}
\label{PcPis}
\|x\|_C \equiv 2 \sqrt{P\wbullet P} = \frac{\sqrt{(R^2 - x^2)^2 - 4 R^2 L^2}}{R}~.
\end{equation}
 
The one-point function $\big\langle \,W(C) O(x)\,\big\rangle 
= \big\langle\, W(C)\, \widehat O(P)\, \big\rangle$ must depend on 
$\|x\|_C$, and must be homogeneous of degree $\Delta$ in it; thus it must necessarily be of the form
\begin{equation}
\label{opW}
\big\langle\, W(C) \,\widehat O(P) \,\big\rangle = \frac{A_O}{(2\pi \|x\|_C)^\Delta}~.
\end{equation} 
The $2\pi$ factor is inserted for convenience and the constant 
$A_O$ is related to the value of the correlator at $x=0$, {\it{i.e.}} at 
$P = P_0 = (\frac R2,\frac R2, \vec 0)$ where $\|x\|_C \to R$, so that
\begin{equation}
\label{opW0}
\big\langle\, W(C) \,\widehat O(P_0)\, \big\rangle = \frac{A_O}{(2\pi R)^\Delta}~.
\end{equation} 

\section{Calculation of $\delta A_{(4)}$ and $\delta A_{(2,2)}$ at two loops}
\label{app:color}
We provide some details for the calculation of the color factor in the amplitude $\delta A_{(4)}$ 
and $\delta A_{(2,2)}$ at two loops. 

\subsection*{$\delta A_{(4)}$ at two loops}
When $\vec{n}=(4)$, the tensor $R_{(4)}$ associated to the chiral operator $O_{(4)}$
can be written as a normalized sum over all permutations of the generators in
$\tr \big(T^{b_1}T^{b_2}T^{b_3}T^{b_4}\big)$, up to cyclic rearrangements, namely
(see also footnote~\ref{footnote:R})
\begin{equation}
R_{(4)}^{\,b_1 b_2 b_3 b_4}
=\frac{1}{4!}\,4\sum_{p\in S_3}\tr \big(T^{b_{p(1)}}T^{b_{p(2)}}T^{b_{p(3)}}T^{b_4}\big)~.
\label{R4}
\end{equation}
Using this, we can easily compute the tree-level amplitude $A_{(4)}\big|_{\mathrm{tree-level}}$
given in (\ref{Antree0}):
\begin{equation}
A_{(4)}\Big|_{\mathrm{tree-level}} \,=\,\frac{g^4}{4N}\, \tr \big(T^{b_1}T^{b_2}T^{b_3} T^{b_4} 
\big) \,R^{b_1b_2b_3b_4}_{(4)}~.
\label{A40}
\end{equation}
Using the explicit form (\ref{R4}), one can realize that 
$ \tr \big(T^{b_1}T^{b_2}T^{b_3} T^{b_4} \big) \,R^{b_1b_2b_3b_4}_{(4)}$ contains six terms
that have three different structures. The first one is
\begin{equation}
\begin{aligned}
\frac{1}{6}\,\tr \big(T^{b_1}T^{b_2}T^{b_3}T^{b_4}\big)\,
&\tr \big(T^{b_1}T^{b_2}T^{b_3}T^{b_4}\big)\\
&\hspace{-0.9cm}=\frac{1}{6}\bigg[\frac{1}{8}\big(d^{b_1b_2e}+\ii f^{b_1b_2e}\big)
\big(d^{b_3b_4e}+\ii f^{b_3b_4e}\big)+\frac{1}{4N}\,\delta^{b_1b_2}\delta^{b_3b_4}\bigg]^2\\
&\hspace{-0.9cm}=\frac{(N^2-1)(N^2+3)}{96 N^2}\phantom{\bigg|}
\end{aligned}
\label{color1}
\end{equation}
where the last equality follows from the group theory identities in Appendix~\ref{app:group}. 
The second type of structure is
\begin{equation}
\begin{aligned}
\frac{1}{6}\,\tr \big(T^{b_1}T^{b_2}T^{b_3}T^{b_4}\big)\,
& \tr \big(T^{b_2}T^{b_1}T^{b_3}T^{b_4}\big) \\
&\hspace{-0.9cm}=\frac{1}{6}\bigg[\frac{(N^2-1)(N^2+3)}{16 N^2}+\ii f^{b_2b_1c}\,
\tr \big(T^{c}T^{b_3}T^{b_4}\big)\,\tr \big(T^{b_1}T^{b_2}T^{b_3}T^{b_4}\big)\bigg]\\
&\hspace{-0.9cm}= -\frac{(N^2-1)(N^2-3)}{96 N^2}\phantom{\bigg|}~.
\end{aligned}
\label{color2}
\end{equation}
Up to relabeling of the indices, we have four such terms. Finally, the third structure is
\begin{align}
\frac{1}{6}\,\tr \big(T^{b_1}T^{b_2}T^{b_3}T^{b_4}\big)\,
& \tr \big(T^{b_3}T^{b_2}T^{b_1}T^{b_4}\big) \notag\\
&\hspace{-0.9cm}=\frac{1}{6}\bigg[-\frac{(N^2-1)(N^2-3)}{16 N^2}+\ii 
f^{b_4b_3c}\,\tr \big(T^{c}T^{b_2}T^{b_1}\big)\,\tr \big(T^{b_1}T^{b_2}T^{b_3}T^{b_4}\big)\bigg]
\notag\\
&\hspace{-0.9cm}=\frac{(N^2-1)(N^4-3N^2+3)}{96 N^2}\phantom{\bigg|}~.
\label{color3}
\end{align}
Summing these contributions and plugging the result in (\ref{A40}), we get
\begin{equation}
A_{(4)}\Big|_{\mathrm{tree-level}} \,=\,g^4\frac{(N^2-1)(N^4-6N^2+18)}{384 N^3}~,
\label{A400}
\end{equation}
which precisely matches the matrix model expression reported in the last-but-one row of Tab.~\ref{tab1}.

Now let us consider the two-loop correction $\delta A_{(4)}\big|_{\mathrm{2-loop}}$. 
{From} (\ref{deltaAn2loopfin}), we have
\begin{align}
\delta A_{(4)}\Big|_{\mathrm{2-loop}}&= -g^4\,\frac{3\, \zeta(3)}{(8 \pi^2)^2}\,
\Big[\frac{g^4}{4N}\,R_{(4)}^{\,b_1b_2b_3 b_4}\,
\tr \big(T^{a_1}T^{a_2}T^{a_3} T^{a_4}\big)\Big]\label{deltaA42loopfin}\\
&\hspace{-0.75cm}
\times \Big[4\,(N^2+1)\,\delta^{b_1a_1}\,\delta^{b_2a_2}\,\delta^{b_3a_3}\,\delta^{b_4a_4}
-2\!
\sum_{p \in S_{3}} \!C_4^{\,b_1b_2 a_{p(1)} a_{p(2)}}
\, \delta^{b_3a_{p(3)}}\,\delta^{b_4 a_4}\,\Big]~.\notag
\end{align}
The first term in the square brackets, which corresponds to the sub-amplitude $I_{(4)}$ associated
to the diagram $\mathbf{(i)}$ of Fig.~\ref{fig:WOn2loop}, is proportional to the tree-level
result (\ref{A400}) and is given by 
\begin{equation}
I_{(4)}= -g^4\,
\frac{\zeta(3)}{(8\pi^2)^2}\,\Big[g^4\frac{(N^2-1)(N^2+1)(N^4-6N^2+18)}{32 N^3}\Big]~.
\label{I4}
\end{equation}
The second term in the square brackets of (\ref{deltaA42loopfin}), corresponding to the
sub-amplitude $J_{(4)}$ associated to the diagram $\mathbf{(j)}$ of Fig.~\ref{fig:WOn2loop}, 
is a bit lengthy to compute, since it is no more proportional to the tree-level expression (\ref{A400}).
However, looking at the explicit form of the tensor $C_4$ which we rewrite here for convenience
\begin{equation}
C_4^{\,b_1b_2a_1a_2}=-\frac{1}{2}\big(\delta^{b_1a_1}\,\delta^{b_2a_2}+\delta^{b_1a_2}\,\delta^{b_2a_1}+\delta^{b_1 b_2}\,\delta^{a_1 a_2}\big)~,
\label{C4app}
\end{equation}
we can realize that
\begin{equation}
\frac{g^4}{4 N}\,R_{(4)}^{\,b_1b_2b_3 b_4}\,\tr \big(T^{a_1}T^{a_2}T^{a_3}
T^{a_4}\big)\,\delta^{b_1a_{p(1)}}\,\delta^{b_2a_{p(2)}}\,\delta^{b_3a_{p(3)}}\delta^{b_4a_4} 
= A_{(4)}\Big|_{\mathrm{tree-level}}
\end{equation}
for any permutation $p\in S_3$, thanks to the symmetry of $R_{(4)}$. Thus, the first two terms
of $C_4$ produce color structures that are proportional to the tree-level one for each permutation $p$.
We can therefore write
\begin{equation}
\begin{aligned}
J_{(4)}&= -g^4\,\frac{3\,\zeta(3)}{(8\pi^2)^2}\,\bigg[ 12\,A_{(4)}\Big|_{\mathrm{tree-level}}\\
&\qquad~~+
\frac{g^4}{4 N}\,R_{(4)}^{\,b_1b_2b_3 b_4}\,\tr \big(T^{a_1}T^{a_2}T^{a_3}
T^{a_4}\big)\sum_{p\in S_3}\,\delta^{b_1b_2}\, \delta^{a_{p(1)}a_{p(2)}} \,\delta^{b_3a_{p(3)}}\,\delta^{b_4 a_4}
\bigg]~.
\end{aligned}
\end{equation}
The last term must be computed explicitly. To do so we use the fact that
\begin{equation}
R_{(4)}^{\,a a b c}= \frac{2N^2-3}{12N}\,\delta^{bc}~,
\end{equation}
so that
\begin{equation}
\begin{aligned}
J_{(4)}&= - g^4\,\frac{3\, \zeta(3)}{(8 \pi^2)^2} \bigg[12\, A_{(4)}\Big|_{\mathrm{tree-level}} \\
&~~~~~~\quad+\frac{g^4}{4 N}\,\frac{2N^2-3}{12N}\,\Big(4\,\tr \big( T^aT^aT^bT^b \big)+ 2
\tr \big( T^aT^bT^aT^b \big)\Big)\bigg] \\ 
&= - g^8\,\frac{3\, \zeta(3)}{(8 \pi^2)^2} \,\frac{1}{4N}\,\bigg[
\frac{(N^2-1)(N^4-6N^2+18)}{8 N^2}+\frac{(N^2-1)(2N^2-3)^2}{24 N^2}\bigg] \\
&=- g^8\,\frac{\zeta(3)}{(8 \pi^2)^2}\,\frac{(N^2-1)(7N^4-30N^2+63)}{32 N^3}~.
\end{aligned}
\end{equation}
Notice that in the large-$N$ limit, $J_{(4)}$ is subleading with respect to $I_{(4)}$. 
Summing the two contributions, we find that 
the total amplitude $\delta A_{(4)}\big|_{\mathrm{2-loops}} $ is
\begin{equation}
\delta A_{(4)}\Big|_{\mathrm{2-loops}} = I_{(4)} + J_{(4)}
=-g^8\frac{\zeta(3)}{(8\pi^2)^2}
\frac{(N^2-1)(N^6+2N^4-18N^2+81)}{32N^3}
\end{equation}
which exactly matches the matrix model expression reported in the last-but-one row of Tab.~(\ref{tab2}).

\subsection*{$\delta A_{(2,2)}$ at two loops}
In a similar way we perform the computation for the other 4-dimensional operator, namely $O_{(2,2)}$, defined by the tensor
\begin{equation}
R_{(2,2)}^{\,b_1 b_2 b_3 b_4}
=\frac{1}{4!}\,4\!\sum_{p\in S_3}\tr \big(T^{b_{p(1)}}T^{b_{p(2)}}\big)\: \tr \big(T^{b_{p(3)}}T^{b_4}\big)=\frac{1}{12}\big(\delta^{b_1 b_2}\delta^{b_3 b_4}+\delta^{b_1 b_3}\delta^{b_2 b_4}+\delta^{b_2 b_3}\delta^{b_1 b_4}\big)~.
\label{R22}
\end{equation} 
Then, from (\ref{Antree0}) the tree-level amplitude:
\begin{equation}
\begin{aligned}
A_{(2,2)}\Big|_{\mathrm{tree-level}} \,&=\,\frac{g^4}{4N}\, \tr \big(T^{b_1}T^{b_2}T^{b_3} T^{b_4} 
\big) \,R^{b_1b_2b_3b_4}_{(2,2)}\\ &=\,\frac{g^4}{4N}\,\frac{1}{12}\left[\,2 \,\tr \big(T^aT^aT^b T^b \big)+\tr \big(T^aT^bT^a T^b \big) \right] \\
&=\,\frac{g^4}{4N}\,\frac{(N^2-1)(2N^2-3)}{48N} ~.
\end{aligned}
\label{A220}
\end{equation}
We can see that this matches the matrix model expression reported in the last row of Tab.~\ref{tab1}.

Let us then consider the two-loop correction $\delta A_{(2,2)}\big|_{\mathrm{2-loop}}$. According to (\ref{deltaAn2loopfin}):
\begin{align}
\delta A_{(2,2)}\Big|_{\mathrm{2-loop}}&= -g^4\,\frac{3\, \zeta(3)}{(8 \pi^2)^2}\,
\Big[\frac{g^4}{4N}\,R_{(2,2)}^{\,b_1b_2b_3 b_4}\,
\tr \big(T^{a_1}T^{a_2}T^{a_3} T^{a_4}\big)\Big]\label{deltaA222loopfin}\\
&\hspace{-0.75cm}
\times \Big[4\,(N^2+1)\,\delta^{b_1a_1}\,\delta^{b_2a_2}\,\delta^{b_3a_3}\,\delta^{b_4a_4}
-2\!
\sum_{p \in S_{3}} \!C_4^{\,b_1b_2 a_{p(1)} a_{p(2)}}
\, \delta^{b_3a_{p(3)}}\,\delta^{b_4 a_4}\,\Big]~.\notag
\end{align}
The first term in the square brackets of the last line, which corresponds to the diagram 
of type $\mathbf{(i)}$ in 
Fig.~\ref{fig:WOn2loop}, is manifestly proportional to the tree-level
result (\ref{A220}) and is given by 
\begin{equation}
I_{(2,2)}= -g^4\,
\frac{\zeta(3)}{(8\pi^2)^2}\,\Big[g^4\frac{(N^2-1)(N^2+1)(2N^2-3)}{16 N^2}\Big]~.
\label{I22}
\end{equation}
The second term of the last line of (\ref{deltaA222loopfin}) corresponds to the
sub-amplitude $J_{(2,2)}$ associated to the diagram of type $\mathbf{(j)}$ in
Fig.~\ref{fig:WOn2loop}. Exploiting the symmetry properties of $C_4$ and $R_{(2,2)}$,
we can immediately write it as 
\begin{equation}
\begin{aligned}
J_{(2,2)}&= -g^4\,\frac{3\,\zeta(3)}{(8\pi^2)^2}\,\bigg[ 12\,A_{(2,2)}\Big|_{\mathrm{tree-level}}\\
&\qquad~~+
\frac{g^4}{4 N}\,R_{(2,2)}^{\,b_1b_2b_3 b_4}\,\tr \big(T^{a_1}T^{a_2}T^{a_3}
T^{a_4}\big)\sum_{p\in S_3}\,\delta^{b_1b_2}\, \delta^{a_{p(1)}a_{p(2)}} \,\delta^{b_3a_{p(3)}}\,\delta^{b_4 a_4}
\bigg]~.
\end{aligned}
\end{equation}
Differently from $J_{(4)}$, the form of
\begin{equation}
R_{(2,2)}^{\,a a b c}= \frac{N^2+1}{12}\,\delta^{bc}
\end{equation}
implies that also $J_{(2,2)}$ is proportional to the tree-level amplitude. Indeed,
\begin{equation}
\begin{aligned}
J_{(2,2)}&= -g^4\,\frac{3\,\zeta(3)}{(8\pi^2)^2}\,\bigg[ 12\,+2\,(N^2+1)\bigg] A_{(2,2)}\Big|_{\mathrm{tree-level}}\\
&=-g^4\,
\frac{\zeta(3)}{(8\pi^2)^2}\,\Big[g^4\frac{(N^2-1)(N^2+7)(2N^2-3)}{32 N^2}\Big]~.
\end{aligned}
\end{equation}
We explicitly notice that in this case both $I_{(2,2)}$ and $J_{(2,2)}$ contribute to the leading order in the large-N limit. In total we get:
\begin{equation}
\delta A_{(2,2)}\Big|_{\mathrm{2-loops}} = I_{(2,2)} + J_{(2,2)}
=-g^8\frac{\zeta(3)}{(8\pi^2)^2}
\frac{3\,(N^2-1)(2N^2-3)(N^2+3)}{32N^2}
\end{equation}
which matches the matrix model expression reported in the last row of Tab.~(\ref{tab2}).

\section{Group theory identities}
\label{app:group}
Here we collect some group theory formulas that are useful to perform explicit calculations and
check our results. We take the generators $T^a$ of SU($N$) to be Hermitean and normalized as
\begin{equation}
\tr \big(T^a T^b\big)=\frac{1}{2}\,\delta^{ab}~,
\label{norm1}
\end{equation}
and define the structure constants $f^{abc}$ by
\begin{equation}
\big[T^a\,,\, T^b \big]=\ii\, f^{abc}\,T^c~,
\label{fabc} 
\end{equation}
and the $d^{abc}$-symbols by
\begin{equation}
\big\{T^a\,,\, T^b \big\}=\frac{1}{N}\,\delta^{ab} + d^{abc}\,T^c~.
\label{dabc}
\end{equation}
Then one has
\begin{align}
\tr \big(T^a T^b T^c\big)&=\frac{1}{4}\big(d^{abc}+\ii \,f^{abc}\big)~,\\
\tr \big(T^a T^b T^c T^d\big)&=\frac{1}{8}\big(d^{abe}+\ii\, f^{abe}\big)
\big(d^{cde}+\ii \,f^{cde}\big)+\frac{1}{4N}\,\delta^{ab}\,\delta^{cd}~,
\end{align}
and
\begin{align}
f^{abe}\,f^{cde}&=\frac{2}{N}\,\big(\delta^{ac}\,\delta^{bd}
+\delta^{ad}\,\delta^{bc}\big)+d^{ace}\,d^{bde}-d^{ade}\,d^{bce}~,\\
d^{abc}\,d^{abd}&=\frac{N^2-4}{N}\,\delta^{dc}~,\\
f^{abc}\,f^{abd}&=N\,\delta^{dc}~,\\
f^{abc}\,d^{abd}&=0~.
\end{align}

\end{appendix}

\providecommand{\href}[2]{#2}\begingroup\raggedright\endgroup


\end{document}